\documentclass[12pt]{article}

\usepackage{ragged2e} 
\usepackage{changepage}

\usepackage{siunitx} 
\usepackage{amsmath}
\usepackage{amssymb}
\usepackage{xfrac} 
\usepackage{bbm}

\usepackage{graphicx}
\usepackage{subcaption}
\usepackage{sidecap}
\usepackage{wrapfig}

\usepackage{caption}
\usepackage{verbatim}
\usepackage[english]{babel}

\usepackage{array}
\usepackage{adjustbox}
\usepackage{float}
\usepackage{enumerate}
\usepackage{multirow}
\usepackage{multicol}
\usepackage[table,xcdraw]{xcolor}
\usepackage[round]{natbib}

\usepackage{csvsimple}
\usepackage{bbm}
\usepackage{pdflscape}
\usepackage{breqn}
\usepackage{lscape}

\usepackage{amsfonts}
\usepackage{latexsym}
 \usepackage[titletoc]{appendix}  
\usepackage[colorlinks=true, citecolor= blue]{hyperref} 
\hypersetup{linkcolor=black} 
\usepackage[all]{hypcap} 

\usepackage[a4paper,top=2cm,bottom=2cm,left=3cm,right=3cm,marginparwidth=1.75cm]{geometry}

\usepackage{float} 
\usepackage{xcolor}
\usepackage{setspace}
\usepackage{hyperref}

\usepackage{authblk}

\newcommand{\shortsmile}[1]{\overset{\scalebox{0.6}[0.6]{$\smile$}}{#1}\mspace{-1mu}}

\title{\vspace{-1cm} \huge \textbf{Distributed lag non-linear models with spatial effect modification using Laplacian P-splines}}
\date{}

\author[1]{Sara Rutten
\footnote{Corresponding author. {\textit{E-mail address}: sara.rutten@uhasselt.be}}}
\author[1,2]{Thomas Neyens}
\author[1]{Elisa Duarte}
\author[3]{Antonio Gasparrini}
\author[1]{Christel Faes}

\affil[1]{Interuniversity Institute for Biostatistics and statistical Bioinformatics (I-BioStat), Data Science Institute (DSI), Hasselt University, Hasselt, Belgium}
\affil[2]{L-BioStat, Department of Public Health and Primary Care, KU Leuven, Leuven, Belgium}
\affil[3]{Environment \& Health Modelling (EHM) Lab, Department of Public Health Environments and Society, London School of Hygiene \& Tropical Medicine, London, United Kingdom}

\begin{document}
\maketitle

\begin{abstract}
Distributed lag non-linear models (DLNMs) are a popular approach to flexibly model the effect of time-delayed exposures. Classical DLNMs specify a common exposure-lag-response relationship across geographical areas. However, this relationship might be altered by an effect modifier that differs between spatial units. Although some methods have been proposed to account for effect modification, their applicability is context-dependent. For example, a meta-analysis can account for heterogeneity between groups, but this technique requires sufficiently large study groups. This limitation is particularly relevant when working with count data, where small numbers of events are often encountered. In this paper, we review existing methods that allow for spatial effect modification for count-based outcomes and propose a Bayesian DLNM alternative method that accounts for the modifier through flexible interaction effects. Through the use of Laplacian P-splines, we provide a computationally fast estimation procedure by avoiding the use of classical Markov Chain Monte Carlo (MCMC) approaches. The performance of the different methods is evaluated through simulation studies. Moreover, the practical applicability of our proposed method is showcased through a data application, containing daily temperature and mortality count data in $87$ Italian cities.
  
  
 \thispagestyle{empty}
\end{abstract}

\newpage
\clearpage\null

\pagenumbering{arabic}

\section{Introduction}
Distributed lag non-linear models (DLNMs), introduced by \cite{gasparrini2010}, have become popular for modelling time-delayed exposure effects in various scientific fields. Through the use of splines, DLNMs provide a powerful framework for quantifying complex exposure-lag-response relationships. \cite{Gasparrini2017} extended the original formulation by replacing low-dimensional unpenalized spline bases with high-dimensional penalized spline bases, allowing for greater flexibility while controlling smoothness. In environmental health contexts, DLNMs are frequently used to smoothly model the delayed effect of environmental exposures such as temperature on health outcomes \citep{Gasparrini2015, Bao2016, Guo2022}. Also in the context of infectious disease modelling, DLNMs have commonly been applied. For example, in \cite{Fajgenblat2024}, they were used to investigate the effectiveness of interventions on the COVID-19 incidence while \cite{Rutten2025} utilized DLNMs to investigate the effect of pollution on COVID-19 incidence. Moreover, in ecological contexts, \cite{Redana2024} applied the DLNM to examine how past temperature exposure influences thermal tolerance in stonefly larvae, while \cite{Xu2024} used these models to investigate drought impacts on vegetation health.

Standard DLNMs assume a common exposure-lag-response relationship across the population, ignoring possible heterogeneity between subgroups. However, factors like age, sex, or socioeconomic status (SES) are likely to modify exposure-health relationships \citep{Yang2012, Rutten2024}. Similarly, species traits, life stage, or habitat conditions might alter the relationship between environmental exposures and ecological responses \citep{Albaladejo-Robles2023, Guezen2025}. A simple approach to take this into account is the use of a stratified analysis, where a different DLNM component is fitted for every pre-defined subgroup of the population \citep{Yang2012, Rutten2024, Duarte2025}. However, this method can be highly inefficient for small subgroups since no information is borrowed across groups. Another solution, named the Bayesian Distributed Lag Interaction Model (BDLIM), has been proposed by \cite{Wilson2017}. However, this method builds on the linear distributed lag models, which allow for a much less flexible relationship than the DLNMs. Moreover, similar to the stratification method, this approach is limited to studying pre-defined subgroups, not allowing for a continuous effect modifier.

In multi-area settings, effect modification by location is a key concern, since area-specific characteristics may modify the exposure–lag–response association. Such heterogeneity cannot be detected when area-specific data are aggregated over the entire study region. Therefore, several extensions of the standard DLNM framework have been proposed. \cite{Gasparrini2013} introduced a two-stage approach in which the DLNM is fitted independently in each geographical area, followed by a meta-analysis. This method can naturally take into account modifying covariate effects through the inclusion of meta-variables. This two-stage method is widely used \citep{TAO2025101653, Gasparrini2015, Mano2024}; however since it requires medium- to large-sized areas at the first stage, it is not well suited for small-area analyses, such as municipality-level time series. Due to the limited information available in small areas, methods that allow information sharing across areas are preferable. A small-area DLNM framework was proposed by \cite{Gasparrini2022}, extending the case time series approach to account for area-specific baseline risks. Within this model, spatial effect modification is accommodated by specifying a linear interaction between the distributed lag non-linear term and the modifier. Similarly, in a Bayesian framework, \cite{Lowe2021} modelled spatially varying  baseline risks using spatial random effects, combined with a linear interaction between the DLNM term and a continuous covariate to study the effect of urbanisation and frequency of water supply shortages. However, these approaches rely on unpenalized DLNMs, which restrict the flexibility of the estimated exposure–lag–response functions. Another small-area DLNM approach for count data was proposed by \cite{Rutten2026}. Their Bayesian DLNM-LPS (DLNM with Laplacian P-splines) framework uses Laplacian P-splines (LPS) \citep{gressani2018} to replace low-dimensional unpenalized spline bases with high-dimensional penalized representations, thereby allowing for greater flexibility in the estimated exposure–lag–response surface. Moreover, in contrast to the case time series approach, this method explicitly quantifies spatial variation in baseline risk by the use of a CAR-type spatial random effect. However, the method assumes a common (spatially invariant) exposure-lag-response relationship, and therefore does not accommodate effect modification by location.

Despite the growing popularity of the DLNMs in spatial statistics, methodological developments that explicitly address spatial effect modification within this framework remain limited. In this paper, we review existing methods and extend the framework of \cite{Rutten2026} to allow for both continuous and categorical effect modifiers of the exposure-lag-response relationship in Poisson or negative binomial regression models. Our extension preserves the advantage of the use of penalized DLNMs and spatial random effects, while further incorporating flexible interactions between the distributed lag non-linear term and (smooth) effect modifiers. To ensure computational efficiency, we use Laplace approximations instead of the traditional Markov chain Monte Carlo (MCMC) approaches \citep{gressani2018, lambert2023, gressani2024, Sumalinab2024}. 

Section~2, reviews two existing approaches that allow for spatial effect modification within the DLNM framework. Section~3 introduces the proposed methodological extension of \cite{Rutten2026}. Section~4 presents a simulation study designed to identify settings in which the different methods are most suitable, considering both small and large areas contexts as well as different forms of effect modification. Section~5 illustrates the proposed approach with a real-world data application.

\section{Review existing methods}
\subsection{Two-stage approach}
In medium to large-scale multi-location studies, the two-stage meta-analysis approach, proposed by \cite{Gasparrini2013}, is a widely used modelling method to analyse count data. In the first stage, a DLNM is fitted separately for each location $j = 1,\dots,J$, typically using unpenalized splines \citep{gasparrini2010} and a Poisson or negative binomial regression model. Let $y_{t,j}$ denote the observed counts at time $t = 1,\dots,T$ and location $j=1,\dots,J$. The effect of a delayed exposure $x$ is modelled through the following mean function:
\begin{align*}
    &\log(\mu_{t,j}) = \beta_{0,j}+\sum_{h=1}^{H}{\beta_{h,j} a_{t,j,h}}+s(x_{t,j}, \ldots , x_{t-L,j} ; \boldsymbol{\theta_j}),
\end{align*}
where $\mu_{t,j}=\mathbb{E}(y_{t,j})$, $a_{t,j,h}$ denote additional covariates used to adjust for confounding and temporal variation in the outcome, and $\beta_{h,j}$ the associated coefficient. Following the formulation of \cite{gasparrini2010}, the DLNM is defined through the smooth cross-basis function $ s(x_{t,j}, \ldots , x_{t-L,j} ; \boldsymbol{\theta_j})$, which can be written as:
\begin{align}
\label{eqn:DLNM}
    s(x_{t,j}, \ldots , x_{t-L,j} ; \boldsymbol{\theta_j})&=\sum_{i=1}^{v_x} \sum_{k=1}^{v_l}\left(\sum_{l=0}^L\widetilde{b}_i(x_{t-l,j})\shortsmile{b}_k(l) \right)\theta_{j,ik},
\end{align}
where $\{\widetilde{b}_i\}_{i=1}^{v_x}$ and $\{\shortsmile{b}_k\}_{k=1}^{v_l}$ denote spline basis functions for the exposure $(x_{t-l,j})$ and lag $(l)$ dimensions, respectively. 
This model can also be written in matrix formulation (see Supplementary Materials Section~1.1). The parameter vector $\boldsymbol{\theta_j}  = (\theta_{j,11},\dots, \theta_{j,1v_{l}},\dots \theta_{j,v_{x}1},\dots, \theta_{j,v_{x}v_{l}})^{\top}$ contains all the area-specific DLNM coefficients. 

In the second stage, a multivariate meta-regression is used to pool the area-specific estimates $\boldsymbol{\hat{\theta}_j}$ i.e.:
\begin{align*}
    \boldsymbol{\hat{\theta}_j} \sim \mathcal{N}(Z_j\boldsymbol{\vartheta},S_j+\Psi), 
\end{align*}
where $Z_j$ contain the area-level effect modifiers and $\boldsymbol{\vartheta}$ are the corresponding meta-regression coefficients. The matrix $S_j$ represents the variance-covariance matrix of $\boldsymbol{\hat{\theta}_j}$, estimated in the area-specific first stage model, while $\Psi$ represents the additional between-location variance-covariance matrix, reflecting the variability of the area-specific estimates that cannot be explained by within-location sampling error $S_j$ or the effect modifiers $Z_j$. 

\subsection{Case time series design}
For small area studies, with effect modifier $z$, \cite{Gasparrini2022} applied the DLNM framework within the case time series design. For conditional Poisson or negative binomial regression, the mean can be modelled through:
\begin{align*}
    &\log(\mu_{t,j}) = \beta_{0,j}+\sum_{h=1}^{H}{\beta_h a_{t,j,h}}+\sum_{i=1}^{v_x} \sum_{k=1}^{v_l}\left(\sum_{l=0}^L\widetilde{b}_i(x_{t-l,j})\shortsmile{b}_k(l) \right)(\theta^{(1)}_{ik}+\theta^{(2)}_{ik}z_j ),
\end{align*}
with $\theta_{ik}^{(1)}$ and $\theta_{ik}^{(2)}$ the regression coefficients of the DLNM main and interaction effect respectively, $\beta_{0,j}$ denoting location-specific intercepts, allowing for different baseline risks, and $z_j$ denoting the location-specific level of the effect modifier. This conditional framework removes the effect of time-invariant covariates on the baseline risk but estimates of $\beta_{0,j}$ are not explicitly estimated. Note that this cross-basis function is again constructed from unpenalized splines, typically using only $2$ or $3$ basis functions to avoid overfitting.

\section{Spatial effect modification in DLNM-LPS}

In this section, we first describe the DLNM–LPS model proposed by \cite{Rutten2026}. We then introduce an extension of this framework to accommodate spatial effect modification. 

Assuming that the counts $y_{t,j}$ follow a Poisson or negative binomial distribution, \cite{Rutten2026} specify the mean structure as:
\begin{align}
\label{eq:DLNM_LPS}
    \log(\mu_{t,j}) = \beta_0+\sum_{h=1}^{H}{\beta_h a_{t,j,h}}+ s(x_{t,j}\ldots x_{t-L,j};\boldsymbol{\theta})+u_j,
\end{align}
where $u_j$ represents an area-specific random effect capturing residual spatial variation in the baseline risk. Different prior distributions can be assumed for $u_j$. The simplest choice is an independent Gaussian random effect  $u_j \stackrel{iid}{\sim} \mathcal{N}(0, \tau^{-1})$, with precision $\tau$. However, in small area analysis, spatially structured priors, such as the intrinsic conditional autoregressive (ICAR), Besag–York–Mollié (BYM) \citep{besag1991} or Leroux \citep{leroux1999} models are often more appropriate, as they explicitly account for spatial dependence between neighboring areas.

Bayesian P-splines \citep{lang2004} are used to model the smooth function $s(x_{t,j}\ldots x_{t-L,j};\boldsymbol{\theta})$, by specifying a B-spline basis in both the exposure and lag dimension and imposing a roughness penalty on the corresponding spline coefficients $\boldsymbol{\theta}$. Hence, the prior on $\boldsymbol{\theta}$ is defined as $(\boldsymbol{\theta}|\boldsymbol{\lambda}) \sim \mathcal{N}(\boldsymbol{0},\mathcal{P}^{-1}(\boldsymbol{\lambda}))$, where $\boldsymbol{\lambda}=(\lambda_x,\lambda_l)^{\top}$ are the smoothing parameters controlling the amount of penalization in the exposure and lag dimensions and $\mathcal{P}(\boldsymbol{\lambda})$ is the corresponding penalty matrix. This matrix is constructed as $\mathcal{P}(\boldsymbol{\lambda})=\lambda_x (S_x\otimes I_{v_l}) + \lambda_l (I_{v_x} \otimes S_l)$ with $S_x=D_{v_x}^{\top}D_{v_x} + \delta I_{v_x}$ and $S_{l}=D_{v_l}^{\top}D_{v_l} + \delta I_{v_l}$ where $D_{v_x}$ and $D_{v_l}$ are difference matrices of order $m$ and $\otimes$ denotes the Kronecker product. The Kronecker-sum structure of $\mathcal{P}(\boldsymbol{\lambda})$ induces separate smoothness penalties in the exposure and lag dimensions by penalizing $m$th-order differences between neighboring spline coefficients. In this paper, we set $m=2$. The constant $\delta$ is a small positive value (e.g.  $\delta=10^{-12}$) introduced to ensure that the penalty matrices are full rank. Following \cite{Gasparrini2017}, an additional penalty can be imposed in the lag dimension, encouraging the effect to approach zero at longer lags. This helps identify the correct lag period, even when the specified maximum lag $L$ substantially exceeds the true maximum lag \citep{Obermeier2015}.

\subsection{DLNM-LPS with effect modification}
In the presence of an effect modifier $z$, the model can be extended to allow for an interaction between $z_j$ and the cross-basis function $s(x_{t,j}, \ldots , x_{t-L,j} ; \boldsymbol{\theta})$. The extension $s(x_{t,j}, \ldots , x_{t-L,j} ; \boldsymbol{\theta}(f(z_j)))$ can then can be formulated as:
\begin{align*}
  s(x_{t,j}, \ldots , x_{t-L,j} ; \boldsymbol{\theta}(f(z_j)))  &=\sum_{i=1}^{v_x} \sum_{k=1}^{v_l}\left(\sum_{l=0}^L\widetilde{b}_i(x_{t-l,j})\shortsmile{b}_k(l) \right)\theta_{ik}(f(z_j)) \\
  &=\sum_{i=1}^{v_x} \sum_{k=1}^{v_l}\left(\sum_{l=0}^L\widetilde{b}_i(x_{t-l,j})\shortsmile{b}_k(l) \right)\left(\theta^{(1)}_{ik}+\sum_{r=1}^{v_z}\theta^{(2)}_{ikr}\tilde{c}_r(z_{j})\right),
\end{align*}
where $\{\tilde{c}_r\}_{r=1}^{v_z}$ denotes a basis for variable $z$. Define the coefficient vectors $\boldsymbol{\theta}^{(1)} = (\theta_{11}^{(1)},\dots, \theta^{(1)}_{1v_{l}},\dots \theta^{(1)}_{v_{x}1},\dots, \theta^{(1)}_{v_{x}v_{l}})^{\top}$ and $\boldsymbol{\theta}^{(2)} = (\theta_{111}^{(2)},\dots, \theta_{1v_l1}^{(2)},\ldots,\theta_{v_xv_l1}^{(2)},\ldots,\theta^{(2)}_{v_{x}v_{l}v_z})^{\top}$ and denote $\boldsymbol{\theta} = (\boldsymbol{\theta}^{(1)\top},\boldsymbol{\theta}^{(2)\top})^\top$. To allow for a flexible continuous interaction, a B-spline basis $\{\tilde{c}_r\}_{r=1}^{v_z}$, excluding an intercept, can be chosen for $z$. Then $(\boldsymbol{\theta}|\boldsymbol{\lambda}) \sim \mathcal{N}(\boldsymbol{0},\mathcal{P}^{-1}(\boldsymbol{\lambda}))$ with $\boldsymbol{\lambda}=(\lambda_x^{(1)},\lambda_x^{(2)},\lambda_l^{(1)},\lambda_l^{(2)},\lambda_{z})^{\top}$ the penalty parameters in the different dimensions and with corresponding penalty matrix
\begin{align*}
     \mathcal{P}(\boldsymbol{\lambda})=\text{bdiag}(&\lambda_{x}^{(1)} (S_x\otimes I_{v_{l}}) + \lambda_{l}^{(1)} (I_{v_x} \otimes S_l), \\
     &\lambda_{x}^{(2)} (I_{v_z} \otimes S_x\otimes I_{v_l}) + \lambda_{l}^{(2)} (I_{v_z} \otimes I_{v_x} \otimes S_l)+\lambda_z(S_z \otimes I_{v_x} \otimes I_{v_l})),
\end{align*}
with $\text{bdiag}(\cdot)$ defining a block-diagonal matrix, $S_z=D_{v_z}^{\top}D_{v_z} + \delta I_{v_z}$ and $D_z$ a second order difference matrix.  The block-diagonal structure of $\mathcal{P}(\boldsymbol{\lambda})$ reflects the separation between the main exposure–lag surface (specified through coefficients $\boldsymbol{\theta}^{(1)}$) and the interaction surface (specified through coefficients $\boldsymbol{\theta}^{(2)}$). The first block enforces smoothness of the baseline in the exposure and lag dimensions, in a similar way as in (2). The second block additionally penalizes roughness along the modifier dimension. 

As a special case of the general formulation above, the interaction can be restricted to be linear in the effect modifier.  This case is obtained by choosing a one-dimensional (i.e. $v_z = 1$) basis $\tilde{c}_r(z) = z$. In this case, no smoothness penalty is required in the modifier dimension, and the penalty matrix reduces to
\begin{align*}
    \mathcal{P}(\boldsymbol{\lambda})=\text{bdiag}(&\lambda_{x}^{(1)} (S_x\otimes I_{v_{l}}) + \lambda_l^{(1)} (I_{v_x} \otimes S_l), \\
    &\lambda_{x}^{(2)} (S_x\otimes I_{v_l}) + \lambda_{l}^{(2)} (I_{v_x} \otimes S_l)),
\end{align*}
with $\boldsymbol{\lambda}=(\lambda_x^{(1)},\lambda_x^{(2)},\lambda_l^{(1)},\lambda_l^{(2)})^{\top}$.

When the effect modifier is categorical with $F\geq 2$ categories, the general interaction model simplifies by representing the covariate $z$ through dummy variables $z_r$, with $r = 1,\ldots,F-1$. In that case, define $v_z = F-1$ and let $\tilde{c}_r(z) = z_r$. In this case, smoothness is enforced only over the exposure and lag dimensions, while no smoothing across categories is imposed. The corresponding penalty matrix is given by 
\begin{align*}
     \mathcal{P}(\boldsymbol{\lambda})=\text{bdiag}(&\lambda_{x}^{(1)} (S_x\otimes I_{v_{l}}) + \lambda_{l}^{(1)} (I_{v_x} \otimes S_l), \\
     &\lambda_{x}^{(2)} (I_{F-1} \otimes S_x\otimes I_{v_l}) + \lambda_{l}^{(2)} (I_{F-1} \otimes I_{v_x} \otimes S_l)).
\end{align*}
with penalty parameters $\boldsymbol{\lambda}=(\lambda_x^{(1)},\lambda_x^{(2)},\lambda_l^{(1)},\lambda_l^{(2)})^{\top}$.

\subsection{Bayesian model formulation}
As the model is defined in a Bayesian framework, priors need to be defined on the coefficients. Denote the coefficients of the main effect of $z$ by $\boldsymbol{\gamma}$. In the presence of a smooth main effect, constructed using a basis of dimension $v_{z_2}$, the effect can be penalized by the matrix $\mathcal{P}_z(\lambda_{z}^{(2)})=\lambda_{z}^{(2)} S_{z_2}$ with $S_{z_2}=D_{v_{z_2}}^{\top}D_{v_{z_2}} + \delta I_{v_{z_2}}$. In that case, denote by $\boldsymbol{\bar{\lambda}}$ the full vector $\boldsymbol{\bar{\lambda}} = (\boldsymbol{\lambda}^\top,\lambda_{z}^{(2)})^\top$ and the matrix $\mathcal{P}(\boldsymbol{\bar{\lambda}})=\text{bdiag}(\mathcal{P}_z(\lambda_{z}^{(2)}), \mathcal{P}(\boldsymbol{\lambda}))$. When opting for a linear main effect, denote $\boldsymbol{\bar{\lambda}} = \boldsymbol{\lambda}$ and $\mathcal{P}(\boldsymbol{\bar{\lambda}})=\text{bdiag}(\zeta, \mathcal{P}(\boldsymbol{\lambda}))$ with $\zeta = 10^{-5}$. Finally, in the case of a categorical effect modifier, denote by $\boldsymbol{\bar{\lambda}} = \boldsymbol{\lambda}$ and $\mathcal{P}(\boldsymbol{\bar{\lambda}})=\text{bdiag}(\zeta I_{F-1}, \mathcal{P}(\boldsymbol{\lambda}))$. In all cases, it is assumed that $(\boldsymbol{\gamma}^\top, \boldsymbol{\theta}^{(1) \top}, \boldsymbol{\theta}^{(2) \top})^\top\sim \mathcal{N}(\boldsymbol{0},\mathcal{P}^{-1}(\boldsymbol{\bar{\lambda}}))$. Following \cite{Rutten2026}, we assume a Gaussian prior for the fixed effect parameters $\boldsymbol{\beta} \sim \mathcal{N}(\boldsymbol{0},{V}^{-1}_{\boldsymbol{\beta}})$ with ${V}_{\boldsymbol{\beta}} = \zeta {I}_{H+1}$. We denote the variance-covariance matrix of $\boldsymbol{u}$ by $G^{-1}$. Note that, depending on the structure that is assumed on $\boldsymbol{u}$, this variance-covariance matrix is conditional on variance and correlation parameters. Denote all variance parameters by $\boldsymbol{\tau}$ and a possible correlation parameter by $\rho$. The conditional precision matrix for $\boldsymbol{\xi}$ is given by $Q = \text{blkdiag}(V_{\boldsymbol{\beta}}, \mathcal{P}(\boldsymbol{\bar{\lambda}}), G)$.

Assuming a Poisson distribution on the counts $y_{t,j}$, the Bayesian model can be written as:
\begin{align*}
            &(\boldsymbol{y}|\boldsymbol{\xi}) \sim \text{Poisson}(\boldsymbol{\mu}) \text{ with } \log(\boldsymbol{\mu}) = H\boldsymbol{\xi}, \\
            & (\boldsymbol{\xi}|\boldsymbol{\bar{\lambda}}, \boldsymbol{\tau}, \rho) \sim \mathcal{N}(\boldsymbol{0},Q^{-1}), \\
             & (\tau_i|\delta_{\tau_i}) \sim \mathcal{G}\left(\frac{\nu}{2},\frac{\nu \delta_{\tau_i}}{2}\right) \text{for all $\tau_i \in \boldsymbol{\tau}$} , \\
             & \delta_{\tau_i} \sim \mathcal{G}(a,b), \\
             &(\lambda_i|\delta_{\lambda_i}) \sim \mathcal{G}\left(\frac{\nu}{2},\frac{\nu \delta_{\lambda_i}}{2}\right) \text{for all $\lambda_i \in \boldsymbol{\bar{\lambda}}$} ,\\
             & \delta_{\lambda_i} \sim \mathcal{G}(a,b), \\
             & \rho \sim \text{Beta}\left(\frac{1}{2}, \frac{1}{2}\right),
\end{align*}
where $H$ denotes the design matrix (see Supplementary Materials Section~1.2) and $\mathcal{G}(a,b)$ denotes a Gamma distribution with mean $a/b$ and variance $a/b^2$. The choice of these priors is based on \cite{jullion2007} where $a=b$ is chosen to be sufficiently small (e.g., $10^{-5}$) and $\nu$ is fixed ($\nu=3$). 

 Note that an alternative version of this model without penalization can easily be obtained by defining $\mathcal{P}(\boldsymbol{\bar{\lambda}}) = \text{diag}(\zeta)$, a diagonal matrix with elements $\zeta$. In that case, the hyperparameters $\boldsymbol{\bar{\lambda}}$, and associated priors, are no longer required. Moreover, natural splines, imposing the constraint of being linear beyond boundary knots \citep{Perperoglou2019}, can then be considered instead of B-splines. However, these models require user-defined specifications regarding the number of knots and placement of the knots, making them less flexible compared to the DLNM-LPS (P-splines) models.
 
 A negative binomial distribution can also be assumed on the counts $y_{t,j}$, accounting for possible overdispersion, i.e $(\boldsymbol{y}|\boldsymbol{\xi}) \sim \text{NegBin}(\boldsymbol{\mu}, \phi) \text{ with } \log(\boldsymbol{\mu}) = H\boldsymbol{\xi}$. In this notation, $\phi$ is assumed the dispersion parameter (i.e. the variance of $y_{t,j}$ is equal to $\mu_{t,j}+\frac{\mu_{t,j}^2}{\phi}$). Note that a Poisson distribution corresponds to $\phi \to \infty$.

 \subsection{Laplace approximation}
To estimate this Bayesian model computationally efficiently, we implement Laplace approximations. The conditional posterior of $\boldsymbol{\xi}$ can be written as  $p(\boldsymbol{\xi} \vert \boldsymbol{\bar{\lambda}}, \boldsymbol{\tau},\rho, \phi; \mathcal{D}) \propto \mathcal{L}(\boldsymbol{\xi},\boldsymbol{\bar{\lambda}},\boldsymbol{\tau}, \rho, \phi;\mathcal{D}) p(\boldsymbol{\xi} \vert \boldsymbol{\bar{\lambda}}, \boldsymbol{\tau}, \rho, \phi)$ with $p(\boldsymbol{\xi} \vert \boldsymbol{\bar{\lambda}}, \boldsymbol{\tau}, \rho, \phi) \propto \exp{ \left(-\frac{1}{2}(\boldsymbol{\xi}'Q_\xi^\lambda \boldsymbol{\xi} \bigl)\right)}$ and $\mathcal{L}(\boldsymbol{\xi},\boldsymbol{\bar{\lambda}},\boldsymbol{\tau}, \rho, \phi;\mathcal{D}) $ the Poisson or negative binomial likelihood. The conditional posterior can be approximated by a multivariate normal distribution $p(\boldsymbol{\xi} \vert \boldsymbol{\bar{\lambda}}, \boldsymbol{\tau},\rho, \phi; \mathcal{D})=\mathcal{N}(\hat{\boldsymbol{\xi}}_\lambda,\hat{\boldsymbol{\Sigma}}_\lambda)$ where $\hat{\boldsymbol{\xi}}_\lambda$ and $\hat{\boldsymbol{\Sigma}}_\lambda$ denote the posterior mode and inverse of the negative Hessian of $\log p(\boldsymbol{\xi} \vert \boldsymbol{\bar{\lambda}}, \boldsymbol{\tau},\rho, \phi; \mathcal{D})$ respectively. Besides, the posterior mode of the hyperparameters $\boldsymbol{\bar{\lambda}}, \boldsymbol{\tau}, \rho, \boldsymbol{\delta}, \phi$ can be obtained similar to \cite{Rutten2026}, who provided the Rcode on Github.

\subsection{Model comparison}
For a continuous effect modifier, we introduced two types of interactions: a linear and a smooth interaction effect. Moreover, the continuous modifier could be categorized or effect modification might not be present. To decide which model to choose in practice, a model comparison criteria, such as the Deviance Information Criterion (DIC) is required \citep{Spiegelhalter2002}. This criterion is well suited for hierarchical models and/or improper priors. If the deviance is defined as $D(\boldsymbol{\xi}) = -2\log(\mathcal{L}(\boldsymbol\xi, \bar{\boldsymbol{\lambda}}, \boldsymbol{\tau},\rho,\phi;\mathcal{D}))$,
the DIC can be formulated as $D(\hat{\boldsymbol{\xi}}_\lambda)+2p_D$.
Following \cite{Rue2009}, the effective number of parameters $p_D$ can be approximated by $p_D \approx n_{\boldsymbol{\xi}}-\text{Trace}(Q\hat{\boldsymbol{\Sigma}}_\lambda)$, with $n_{\boldsymbol{\xi}}$ the number of parameters in $\boldsymbol{\xi}, \hat{\boldsymbol{\Sigma}}_\lambda$ the posterior covariance matrix and $Q$ the prior precision matrix of $\boldsymbol{\xi}$.

\section{Simulation study}
\subsection{Simulation set-up}
To evaluate the model and estimation performance, we performed a simulation study. Different settings were considered, varying in population size and smoothness of the effect modification.

\subsubsection*{Simulating data}
In each setting, the time-varying predictor $x$ was the daily summer temperature, i.e. June to September, in the $73$ neighbourhoods of Barcelona, during the period 2007-2016 \citep{Quijal-Zamorano2024}. Hence, for each neighbourhood, $1220$ observations were available ($t = 1,\ldots,1220$), which we standardized over a range $0-10$. An area-specific outcome time series $y_{t,j}$ was simulated from a Poisson distribution with mean:
\begin{align*}
    \log(\mu_{t,j}) = \beta_0 + \beta_1 z_j +\sum_{l=0}^{8}f_j\cdot w_j(x_{t-l,j},l, z_j)+u_j+\log(\text{pop\_size}_j),
\end{align*}
with $\beta_1 = -1$ and $u_j$ simulated from a Leroux distribution with $\rho = 0.95$ and $\sigma^2 = 0.2$ to allow for spatially correlated heterogeneity in the baseline risk. 

\begin{figure}
    \centering
    \includegraphics[width=1\linewidth]{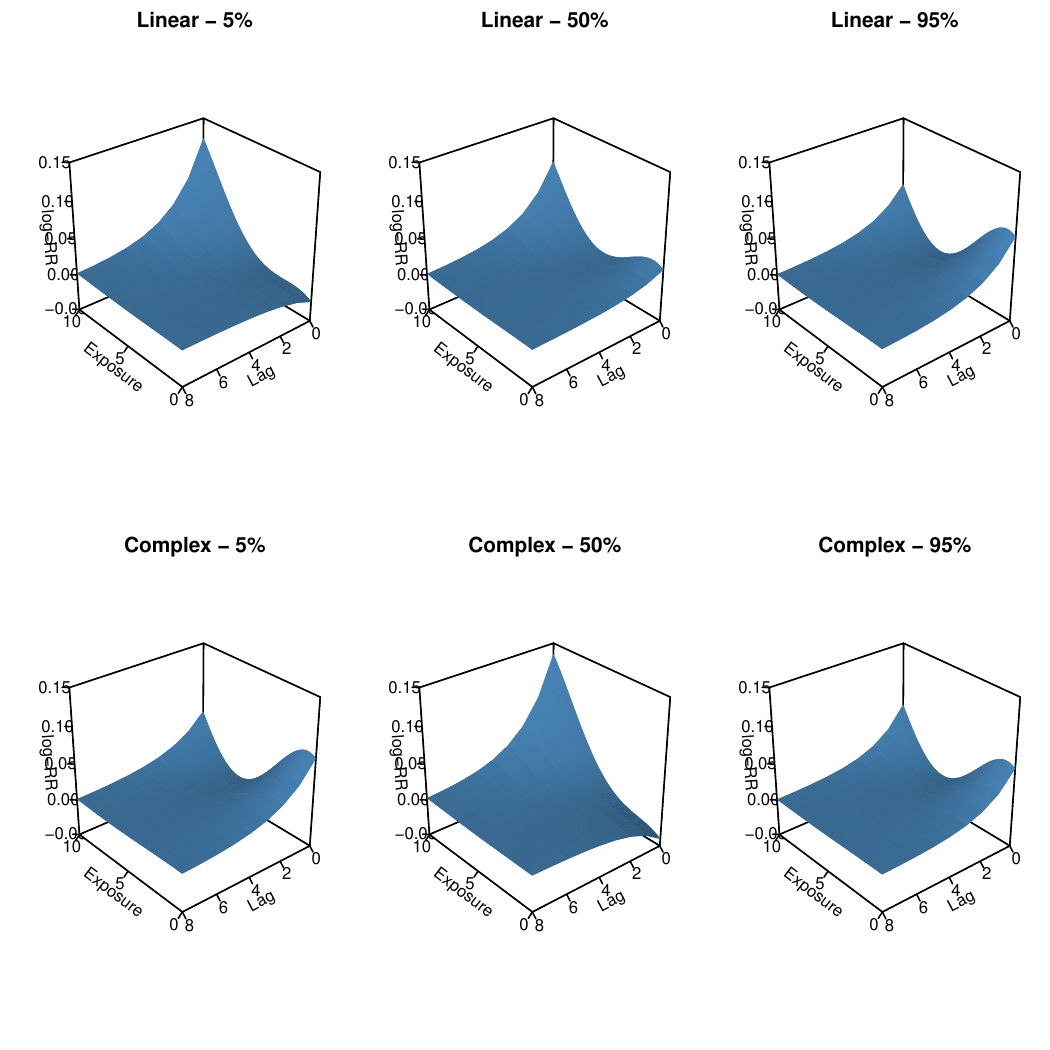}
    \caption{Exposure-lag-response relationship for three different levels of the effect modifier $z$ ($5\%, 50\%$ and $95\%$ percentile of $z$) for the linear case (top) and the complex case (bottom).}
    \label{fig:simulation_scenario}
\end{figure}

The function $f_j\cdot w_j(x_{t-l,j},l, z_j)$ defines the location-specific exposure-lag-response relationship. Two different settings were considered for this function, depending on the complexity of the interaction with the effect modifier $z$ (i.e. linear or complex). In the linear setting, we defined the covariate $z_j \sim \mathcal{N}(0,0.4^2)$, such that $z_j$ falls approximately within the range of $-1$ to $1$. The function $f_j\cdot w_j(x_{t-l,j},l, z_j)$ is then defined such that the log-RR at any exposure level varies linearly with $z$. A 3D figure of the exposure-lag-response relationship is shown in Figure \ref{fig:simulation_scenario} (top) for three different levels of the modifying variable $z$ ($5\%$, $50\%$ and $95\%$ percentile). In the complex setting, $z_j$ was simulated from the same distribution but the function $f_j\cdot w_j(x_{t-l,j},l, z_j)$ now depends on $z$ in a more complex way, through sine and cosine functions. A 3D figure of the exposure-lag-response relationship is again shown in Figure \ref{fig:simulation_scenario} (bottom). More information on the exact specification of these surfaces can be found in the Supplementary Materials. In Figure \ref{fig:lag_effect_simulation}, the lag-response relationship is shown for the exposure $x=8$ and three different levels ($5\%$, $50\%$ and $95\%$ percentile) of the effect modifier $z$.

The intercept $\beta_0$ was set such that the rate $\mu_{t,j}$ corresponds to a realistic mortality rate for Europe, around $0.03$ daily deaths per $1000$ inhabitants \citep{statista_mortality}, resulting in $\beta_0 = 10.5$ and $\beta_0 = 8$ in the linear and complex setting of effect modification respectively.


\begin{figure}
    \centering
    \includegraphics[width=1\linewidth]{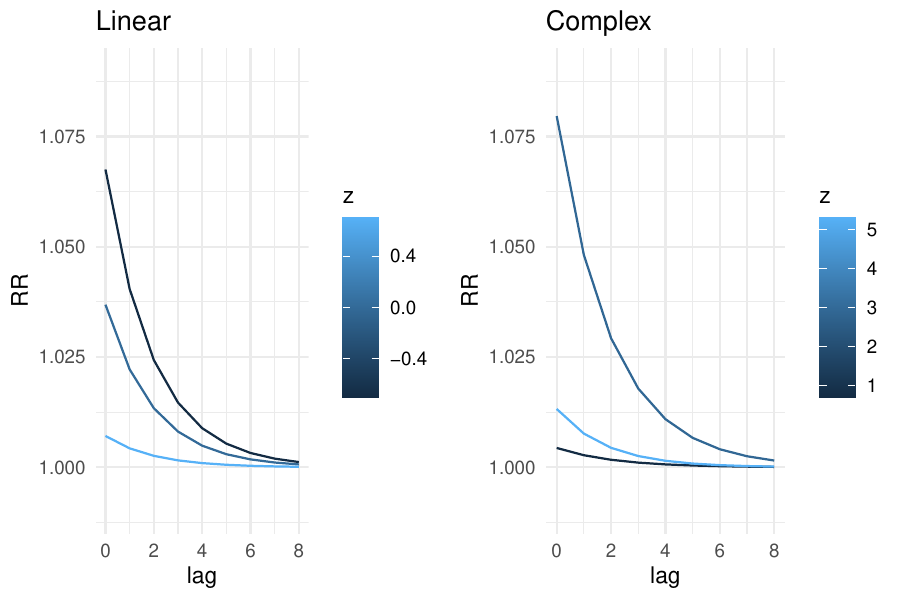}
    \caption{Lag-response relationship for an exposure of $x=8$ and three different levels of the effect modifier $z$ ($5\%, 50\%$ and $95\%$ percentile of $z$) for both linear case (left) and complex case (right).}
    \label{fig:lag_effect_simulation}
\end{figure}
To assess the performance of the proposed and existing methods under both small- and large-area conditions, a neighbourhood-specific population size (pop\_size) was simulated from a lognormal distribution under two distinct settings. In the first setting (Area~1), the log scale mean was set equal to $\log(12000)$ with standard deviation $\sigma = 1.8$, resulting in population sizes ranging from $200$ to $600000$, which is representative for areas comparable in size to Belgian municipalities. In the second setting (Area~2), the log scale mean was set to $\log(6000000)$ with $\sigma = 0.9$, resulting in population sizes ranging from $750,000$ to $42,000,000$, comparable to areas of the size of U.S. states. 

\subsubsection*{Fitted models}
Combining the two different types of effect modification with the two area settings, resulted in four simulation scenarios. In each scenario, we included a Leroux structured random effect and a linear main effect of $z$. For the exposure–lag–response surface, we used P-splines with $8$ degrees of freedom (df), before imposing constraints, in both dimensions of the DLNM cross-basis. We then compared three variants of the proposed DLNM-LPS methodology, differing in how the interaction with the effect modifier was specified: a linear interaction (LPS linear), a smooth interaction with $5$ df (LPS smooth), and a binary interaction based on a median split of $z$ (LPS binary). We included an additional penalty in the lag dimension, encouraging the effect to shrink to zero for longer lags. This penalty matrix was constructed as $S_{l_2} = P_{v_l}+\delta I_{v_l}$ where $P_{v_l}$ is a diagonal matrix with weights given by $p_k = k^2$ with $k = 0, \ldots, (v_l-1)$. 

To benchmark the proposed interaction models against existing approaches, we compared the performance of these three models with that of the DLNM-LPS model without effect modification, proposed by \cite{Rutten2026}, keeping all other model specifications the same. In addition, we evaluated our method against a similar unpenalized Laplace approach, based on natural splines, in order to assess the benefits of penalization in the presence of effect modification. For this comparison, we used splines with $2$ knots in the exposure dimension (placed at the$10\%$ and $90\%$ quantile) and $2$ knots in the lag dimension, equally spaced on the log-scale. Both a linear and a smooth interaction were again considered. 

To assess the impact of the estimation framework and further benchmark the proposed Bayesian approach against commonly used alternatives, we compared the proposed models with maximum-likelihood-based implementations in the \texttt{bam} package in R. This package allows for the specification of a penalized DLNM structure, but does not straightforwardly allow for the use of a Leroux random effect. Therefore, we specified an unstructured random effect. We fitted both the smooth and linear interaction penalized DLNM structure in \texttt{bam}. In addition, we compared our methods with the case time series (CTS) model of \cite{Gasparrini2022}, using the same unpenalized natural spline cross-basis as specified before. Finally, we conducted two-stage meta-analyses \citep{Gasparrini2013}, again using the unpenalized DLNM cross-basis specification and including the modifier as covariate, specified either linearly or smoothly using natural splines with $3$ knots placed on the $10\%, 50\%$ and $90\%$ quantile.

\subsubsection*{Model performance}
The model performance was evaluated through the coverage and root mean squared error (RMSE) of the lag-specific and the overall (cumulative over lag $0-8$) log relative risk (RR) in every region:
\begin{align*}
   \log\bigl(RR_{x,x_0,z}(l)\bigl) &=\sum_{i=1}^{v_x}\sum_{k=1}^{v_l}{\bigl(\widetilde{b}_i(x)-\widetilde{b}_i(x_0)\bigl)\shortsmile{b}_k(l)\hat{\theta}_{ikz}}, 
\end{align*}
and
\begin{align*}
      \log\bigl(RR_{x,x_0,z}^{\text{overall}}\bigl) &=\sum_{i=1}^{v_x}\sum_{k=1}^{v_l}\biggl(\sum_{l=0}^{L}{\bigl(\widetilde{b}_i(x)-\widetilde{b}_i(x_0)\bigl)\shortsmile{b}_k(l)\biggl)\hat{\theta}_{ikz}}, 
\end{align*}
with $\theta_{ikz}$ the coefficients $\theta_{ik}$ modified by covariate $z$. These quantities represent the risk at exposure levels $x$, compared to a reference exposure level $x_0 = 5$, for an area with the level of the effect modifier equal to $z$. They are calculated over a grid of exposure levels $x$ ($x = 0,0.25,\ldots 9.75,10$) and afterwards summarized using the RMSE and coverage (see Supplementary Materials).

\subsection{Simulation results}
We ran $250$ simulations in each scenario. Because of the high computational cost, we performed the simulations on the Flemish Super Computer (VSC). Note that the \texttt{bam} method was timed out if convergence was not reached after $20$ minutes, to avoid excessive computation time. The results, averaged over the runs that are successfully completed for all methods, are shown in Table \ref{tab:simulation_study_linear} and Table \ref{tab:simulation_study_smooth}. To present a realistic view of the computational cost on a laptop, the reported time is the time obtained from a device with an AMD Ryzen(TM) 5 PRO 7530U processor (base frequency 2.00GHz), having six cores (12 threads) and 16GB of RAM, but only averaged over $10$ runs. 

The results in Table \ref{tab:simulation_study_linear} show that, in case of a simple linear effect modifier, both the LPS method allowing for a smooth interaction and the LPS method specifying a linear interaction show satisfying coverage for the small-area and large-area settings. The RMSE is slightly lower for the linear LPS method, although differences are small. Both methods outperform the method assuming a common exposure-lag-response relationship or a binary modifier based on median-split, especially in the presence of large areas. Furthermore, it can be seen that the Laplace methods with an unpenalized DLNM specification, result in an increase in RMSE compared to the penalized methods, especially when allowing for a smooth interaction. Moreover, the CTS method of \cite{Gasparrini2022} performs similar to the unpenalized Laplace method that allows for a linear interaction. Looking at the meta-analysis method introduced by \cite{Gasparrini2013}, we can see an increased RMSE compared to the LPS methods, especially for small areas. This is not unexpected since fitting a separate DLNM in every area can be highly unstable in this setting. Lastly, we fitted a model similar to LPS smooth and LPS linear in the \texttt{bam} package. We can see that this method shows slight undercoverage, especially in the small area setting, but RMSEs are comparable with the LPS methods. In terms of computation time, the penalized methods are slower than the unpenalized ones, both for LPS and \texttt{bam}, but computation time remains rather low. The LPS methods show superior computation times compared to the \texttt{bam} method. Note that for $16\%$ and $6\%$ of the simulations in the small and large area settings, respectively, the model allowing for a smooth interaction fitted in \texttt{bam} is stopped because the computation time exceeds $20$ minutes, in contrast to the LPS method.

Looking at the results of the complex effect modification setting in Table \ref{tab:simulation_study_smooth}, similar observations can be made. However, as expected, the methods that do not allow for a smooth interaction between the covariate $z$ and the DLNM cross-basis, are no longer flexible enough to capture the underlying trends. The LPS linear, CTS method and the linear \texttt{bam} method show a drastic increase in RMSE and decrease in coverage compared to the smooth LPS and the smooth \texttt{bam} method. In contrast, it can be noted that the two meta-analyses methods perform relatively similarly to each other for large areas, which is not entirely surprising since a random effects meta-analysis accounts for variability in true effect sizes. 

We performed a similar simulation assuming a negative binomial model (with dispersion parameter equal to $5$ i.e. $\text{Var}(y_{t,j}) = \mu_{t,j}+\frac{\mu_{t,j}^2}{5}$). Similar conclusions can be drawn from these results, which are given in the Supplementary Materials (Table S1 and Table S2).

\begin{table}
\caption{Simulation results for linear effect modification: RMSE of lag-specific RR (RMSE RR), RMSE of cumulative RR, over lag $0-8$, (RMSE RR overall), coverage of lag-specific RR (cov RR) and coverage of cumulative RR (cov RR overall). The computation time is reported in seconds (averaged over $10$ runs), as well as the proportion of failed (or timed-out) simulations (over 250 runs).}
\label{tab:simulation_study_linear}
\centering
\begin{adjustbox}{width=\linewidth} 
\begin{tabular}{lllllcccccc}
\hline
 \textbf{Area} &  \textbf{Method} & \textbf{Modification} &\textbf{Penalty} & \textbf{RE}& \textbf{failed} & \textbf{time} & \textbf{cov RR} &\textbf{cov RR} & \textbf{RMSE RR} &\textbf{RMSE RR}  \\ 
  &   &  &  & &  &  &  &\textbf{overall} &  &\textbf{overall}  \\ \hline
 \multirow{11}{*}{Small} &  Laplace (LPS) & smooth & yes & Leroux & $0$ & $74.04$ & $0.9777$ & $0.9699$ & $0.0110$ & $0.0457$ \\
                          &  Laplace (LPS) & linear & yes & Leroux & $0$ & $24.18$ & $0.9767$ & $0.9682$ & $0.0104$ & $0.0415$  \\
                          &  Laplace (LPS) & binary & yes & Leroux & $0$ & $20.10$ & $0.9674$  & $0.8786$ & $0.0118$ & $0.0504$ \\
                          &   Laplace (LPS) & none & yes & Leroux & $0$ & $11.92$ & $0.9284$ & $0.6842$ & $0.0109$& $0.0537$  \\
                          &  Laplace & smooth & no & Leroux & $0$ & $12.07$ &  $0.9380$ & $0.9411$ & $0.0721$ & $0.2025$ \\
                          &  Laplace & linear & no & Leroux & $0$ & $4.12$ & $0.9419$  & $0.9479$ & $0.0202$ & $0.0628$ \\
                          &  BAM & smooth & yes & ind. & $0.16$ & $407.01$ & $0.9070$ & $0.8868$ & $0.0099$ & $0.0449$ \\
                          &  BAM & linear & yes & ind. & $0$ & $43.00$ & $0.8708$ & $0.8286$ & $0.0099$ & $0.0465$ \\
                          &  CTS & linear & no & $\cdot$ & $0$  & $0.31$ & $0.9482$ & $0.9511$ & $0.0202$ & $0.0628$ \\
                         &  Meta &  smooth & no & $\cdot$ & $0$ & $35.04$ & $0.9696$ & $0.9616$ & $0.0548$ & $0.1796$ \\
                          &  Meta & linear & no & $\cdot$ & $0$ & $27.84$ & $0.9727$ & $0.9681$ & $0.0304$ & $0.1049$ \\

\hline
  \multirow{11}{*}{Large} & Laplace (LPS) & smooth & yes & Leroux & $0$  & $72.76$ & $0.9595$ & $0.9419$ & $0.0028$ & $0.0115$ \\
                          &  Laplace (LPS)& linear & yes & Leroux & $0$ & $23.63$ & $0.9545$ & $0.9280$ & $0.0024$ & $0.0101$  \\
                          &  Laplace (LPS) & binary & yes & Leroux & $0$ & $19.46$ & $0.7399$ & $0.2847$ & $0.0049$ & $0.0277$ \\
                          &  Laplace (LPS)& none & yes & Leroux & $0$ & $11.10$ & $0.5717$ & $0.1484$ & $0.0075$ & $0.0448$ \\
                          &  Laplace & smooth & no & Leroux & $0$ & $9.20$ & $0.9008$ & $0.7389$ & $0.0039$ & $0.0174$ \\
                          &  Laplace & linear & no & Leroux & $0$ & $3.74$ & $0.8662$ & $0.6129$ & $0.0028$ & $0.0153$ \\
                          &  BAM & smooth & yes & ind. & $0.06$ & $294.44$ & $0.9389$ & $0.9065$ & $0.0024$ & $0.0113$ \\
                          &  BAM & linear & yes & ind. & $0$ & $147.13$ & $0.9306$ & $0.8835$ & $0.0023$ & $0.0115$ \\
                          &  CTS & linear & no & $\cdot$ & $0$ & $0.34$ & $0.8759$ &  $0.6427$ & $0.0028$ & $0.0153$ \\
                          &  Meta &  smooth & no & $\cdot$ & $0$ & $40.46$ & $0.9576$ & $0.8806$ & $0.0040$ & $0.0178$ \\
                          &  Meta & linear & no & $\cdot$ & $0$ & $30.96$ & $0.9563$ & $0.8588$ & $0.0033$ & $0.0164$ \\

\hline

\end{tabular}
\end{adjustbox}
\end{table}

\begin{table}
\caption{Simulation results for smooth effect modification: RMSE of lag-specific RR (RMSE RR), RMSE of cumulative RR, over lag $0-8$, (RMSE RR overall), coverage of lag-specific RR (cov RR) and coverage of cumulative RR (cov RR overall). The computation time is reported in seconds (averaged over $10$ runs), as well as the proportion of failed (or timed-out) simulations (over 250 runs).}
\label{tab:simulation_study_smooth}
\centering
\begin{adjustbox}{width=\linewidth} 
\begin{tabular}{lllllcccccc}
\hline
 \textbf{Area} &  \textbf{Method} & \textbf{Modification} &\textbf{Penalty} & \textbf{RE}& \textbf{failed} & \textbf{time} & \textbf{cov RR} &\textbf{cov RR} & \textbf{RMSE RR} &\textbf{RMSE RR}  \\ 
  &   &  &  & &  &  &  &\textbf{overall} &  &\textbf{overall}  \\ \hline
 \multirow{11}{*}{Small} & Laplace (LPS) & smooth & yes & Leroux & $0$ & $81.47$ & $0.9826$ & $0.9606$ & $0.0133$ & $0.0618$ \\
                          &  Laplace (LPS) & linear & yes & Leroux & $0$ & $25.99$ & $0.9522$ & $0.7868$ & $0.0264$ & $0.1240$  \\
                         &  Laplace (LPS) & binary & yes & Leroux & $0$ & $21.61$ & $0.9006$ & $0.6308$ & $0.0179$ & $0.0915$ \\
                         &  Laplace (LPS)& none & yes & Leroux & $0$ & $12.57$ & $0.8313$ & $0.3992$ & $0.0168$ & $0.0938$ \\
                          &  Laplace & smooth & no & Leroux & $0$ & $16.97$ & $0.9443$ & $0.9347$ & $0.1855$ & $0.7249$ \\
                          &  Laplace & linear & no & Leroux & $0$ & $4.37$ & $0.8988$ & $0.6956$ & $0.0325$ & $0.1339$ \\
                          & BAM & smooth & yes & ind. & $0.12$ & $328.49$ & $0.9333$ & $0.8780$ & $0.0119$ & $0.0596$ \\
                          &  BAM & linear & yes & ind. & $0$ & $137.17$ & $0.8097$ & $0.5861$ & $0.0202$ & $0.1141$ \\
                          &  CTS & linear & no & $\cdot$ & $0$ & $0.46$ & $0.9084$ & $0.7254$ & $0.0325$ & $0.1339$ \\
                          &  Meta &  smooth & no & $\cdot$ & $0.02$ & $35.04$ & $0.9687$ & $0.9480$ & $0.1732$ & $0.4909$ \\
                          &  Meta & linear & no & $\cdot$ & $0.02$ & $28.03$ & $0.9576$ & $0.8858$ & $0.1051$ & $0.2537$ \\

\hline
 \multirow{11}{*}{Large} & Laplace (LPS)& smooth & yes & Leroux  & $0$ & $80.01$ & $0.9583$ & $0.8910$ & $0.0040$ & $0.0170$ \\
                          & Laplace (LPS) & linear & yes & Leroux  & $0$& $25.48$ & $0.6181$ & $0.1745$ & $0.0175$ & $0.1044$  \\
                          &  Laplace (LPS) & binary & yes & Leroux  & $0$ & $20.71$ & $0.5644$ & $0.1940$ & $0.0132$ & $0.0789$ \\
                          & Laplace (LPS) & none & yes & Leroux &  $0$ & $12.41$ & $0.3820$ & $0.0698$ & $0.0144$ & $0.0878$ \\
                          &  Laplace & smooth & no & Leroux  & $0$& $11.10$ & $0.9043$ & $0.7192$ & $0.0089$ & $0.0297$ \\
                          &  Laplace & linear & no & Leroux  & $0$& $4.19$ & $0.3964$ & $0.0916$ & $0.0177$ & $0.1069$ \\
                          & BAM & smooth & yes & ind. & $0.03$ & $260.13$ & $0.9277$ & $0.8778$ & $0.0035$ & $0.0171$ \\
                          &  BAM & linear & yes & ind.  & $0$& $28.39$ & $0.5527$ & $0.1465$ & $0.0183$ & $0.1091$ \\
                          &  CTS & linear & no & $\cdot$ & $0$ & $0.49$ & $0.4036$ & $0.0983$ & $0.0177$ & $0.1068$ \\
                          &  Meta &  smooth & no & $\cdot$ & $0$ & $43.06$ & $0.9554$ & $0.9099$ & $0.0071$ & $0.0253$ \\
                          &  Meta & linear & no & $\cdot$ & $0$ & $30.98$ & $0.9432$ & $0.9512$ & $0.0050$ & $0.0221$ \\

\hline

\end{tabular}
\end{adjustbox}
\end{table}

\section{Data application}
We demonstrate our method on a data set with daily average temperature and mortality counts in $87$ Italian cities between $2011$ and $2021$ \citep{Masselot2025}. A DLNM model was fitted using $8$ degrees of freedom in both the lag and exposure dimension, before imposing constraints, and a maximum lag of $L = 21$ was defined. The inclusion of an additional penalty in the lag dimension again encouraged shrinkage of the effect to zero for longer lags. We compared several different models, all of them including day of the week as a factor, date as P-splines with $7$ df per year, the proportion of people aged over $65$ and an independent city-specific random effect to account for unobserved baseline mortality differences. The population size was included as an offset. Moreover, the city-specific index of deprivation is available in the data set. Since previous studies suggest socioeconomic context may modify temperature-mortality associations, we aimed to investigate its impact on the exposure-response relationship \citep{Bakhtsiyarava2023, Son2019}. Therefore, we specified a LPS model that allows for effect modification through a linear interaction or a smooth interaction ($5$ df). Both models include the deprivation index as a penalized main effect using P-splines with $10$ df. A map of the analysed cities, coloured by deprivation index, can be found in the Supplementary Materials (Figure S1). Furthermore, we fitted a model that includes deprivation index as a binary indicator (median-split) and therefore also includes a categorical interaction. Finally, using the implementation of \cite{Rutten2026}, we compared the models to a LPS model with a common exposure-lag-response surface, where the main effect of deprivation is again included through P-splines with $10$ df. 

Note that we fitted all models for both the Poisson and negative binomial likelihood. Moreover, note that a DLNM is often fitted on the absolute temperature values. Since the temperature range can be substantially different in the northern and southern parts of Italy, fitting a DLNM on the city-specific temperature percentiles might be more appropriate as compared to on the absolute temperatures \citep{Masselot2025}. Therefore, we replicated the model fitting procedure using the temperature percentiles. All models were compared in terms of DIC. The results, reporting $\Delta\text{DIC}=\text{DIC}_i - \min(\text{DIC})$, can be found in Table \ref{tab:DIC_italy}.

\begin{table}
    \centering
        \caption{$\Delta\text{DIC}$, i.e. $\text{DIC}_i - \min(\text{DIC})$, values of the different models fitted for the data set on mortality in 83 Italian cities, for the Poisson and negative binomial model (NB), using both absolute temperature and temperature percentiles.}
    \begin{adjustbox}{width=\textwidth}
    \begin{tabular}{rrrrrr}
        \hline 
        & & LPS smooth & LPS linear & LPS binary & LPS common \\
        \hline
        \multirow{2}{*}{Poisson} & temperature & $2434$ & $2571$ & $2776$ & $3051$ \\
        & percentiles & $1983$ & $2092$ & $2190$ & $2600$ \\
        \hline
        \multirow{2}{*}{NB} & temperature  & $357$ & $461$ & $636$  & $832$ \\
        & percentiles  & $0$ & $74$ & $165$ & $452$  \\
        \hline
    \end{tabular}
    \end{adjustbox}
    \label{tab:DIC_italy}
\end{table}

The table shows lower DIC values for the negative binomial models, compared to the Poisson models. Moreover, fitting a DLNM on the temperature percentiles instead of the absolute values, decreases the DIC even further. Finally, the models allowing for the most flexible interaction with the deprivation index (LPS smooth) have lower DIC values. In summary, the results suggest that the best-fitting model is the model allowing for a smooth modification of deprivation, using a negative binomial likelihood and a DLNM specification based on the temperature percentiles. 

\begin{figure}
    \centering
    \includegraphics[width=0.7\linewidth]{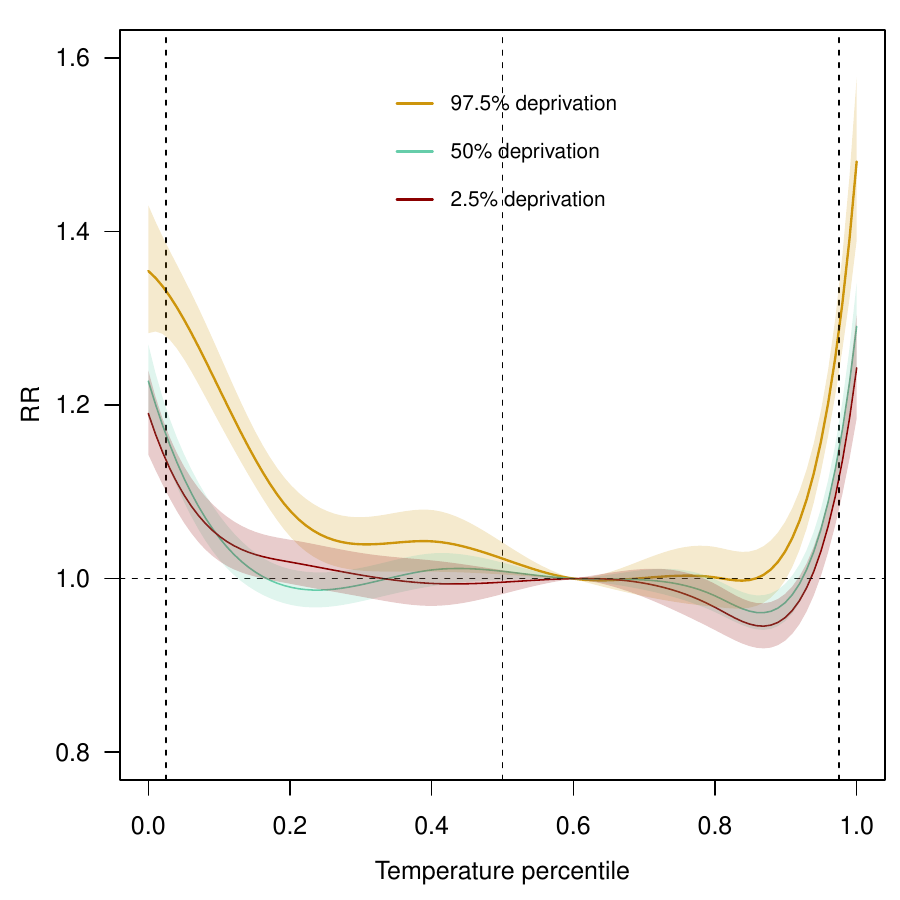}
    \caption{Exposure-response relationship represented by the overall cumulative RR for $3$ deprivation levels, cumulated over 22 days, compared to a temperature percentile of $0.6$. The vertical lines represent the $2.5\%, 50\%$ and $97.5\%$ quantiles of temperature percentiles. The shaded bands indicate the $95\%$ credible intervals.}
    \label{fig:RR_by_cov_smooth}
\end{figure}

\begin{figure}
    \centering
    \includegraphics[width=1\linewidth]{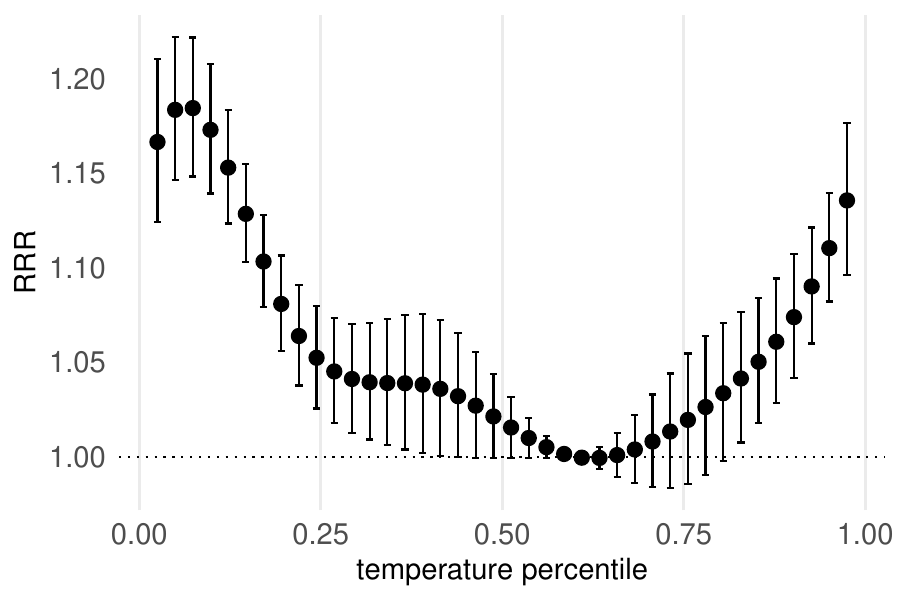}
    \caption{RRR comparing the overall cumulative RR at $90$th and $10$th deprivation percentile for different temperature percentiles, using a reference temperature percentile of $0.6$. The bands are the associated credible intervals.}
    \label{fig:RRR}
\end{figure}

Figure \ref{fig:RR_by_cov_smooth} shows the results of the overall cumulative RR (over $22$ days), compared to a reference temperature percentile of $0.6$, for three different deprivation levels ($2.5\%, 50\%$ and $97.5\%$ quantiles). Similar figures showing the lag-response relationship for a $95\%$ temperature quantile and the estimated overall cumulative RR versus deprivation, for three different temperature quantiles ($2.5\%, 50\%$ and $97.5\%$ quantiles), can be found in the Supplementary Materials (Figure S2 and Figure S3 respectively).  These results suggest that the estimated RRs are higher for the highest deprived cities, especially for the temperature extremes. Following \cite{Achebak2025}, we defined the relative risk ratio (RRR) comparing the risk at the $90$th and $10$th deprivation percentile, by $\text{RRR} = \exp(\log(\text{RR}_{90\text{th deprivation percentile}}) - \log(\text{RR}_{10\text{th deprivation percentile}}))$. The RRRs shown in Figure \ref{fig:RRR} confirm that for the temperature extremes, a significant increase in RR is found for the $90$th deprivation percentile, compared to the $10$th deprivation percentile. It is important to emphasize that this association cannot be interpreted as causal, since it could reflect the influence of unmeasured confounders. A 3D plot and contour plot showing the overall deprivation-temperature-mortality relationship can be found in the Supplementary Materials (Figure S4 and Figure S5). Moreover, a plot of the city-specific random effects is given there (Figure S6).

An important advantage of the Bayesian framework is that it allows direct computation of exceedance probabilities for quantities of interest. In this specific application, we calculated the posterior probability that the RR exceeds $1$, for the different temperature percentiles and deprivation indices. For low temperature percentiles, a clear deprivation association can be seen i.e. cities with a lower deprivation index require more extreme temperature percentiles to attain high exceedance probabilities (Figure S7 in the Supplementary Materials). Although the trend is less pronounced at higher temperatures, exceedance probabilities tend to be larger for above-median deprivation levels, particularly around the 92.5th percentile (Figure S8 in the Supplementary Materials).

\begin{figure}
    \centering
    \includegraphics[width=\linewidth]{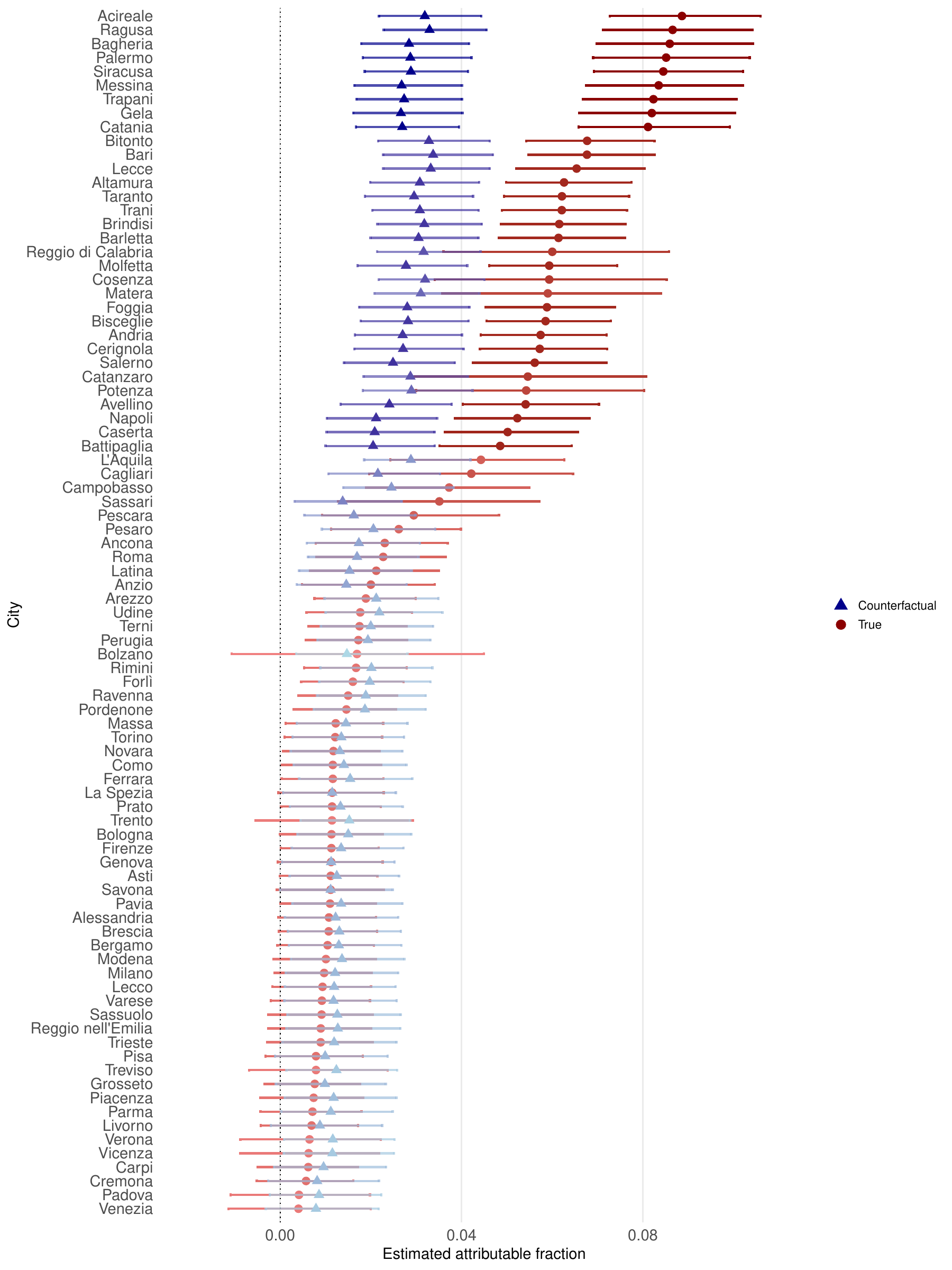}
    \caption{True (red dots) and counterfactual (blue triangle) attributable fraction in $2021$ for the $87$ Italian cities if the deprivation index would be equal to the median for all cities. The darker the colour, the higher the deprivation index.}
    \label{fig:af_smooth_counterfactual}
\end{figure}

Another interesting concept is the attributable fraction (AF) i.e. the percentage of deaths that can be attributed to temperatures differing from the $0.6$ reference temperature percentile. This concept has been formulated in the context of DLNMs \citep{Gasparrini2014, Rutten2026}. Two different perspectives have been introduced: the backward perspective, defining the current risk obtained from a past exposure, and the forward perspective, representing the future risk obtained from a current exposure. Adopting the backward perspective, we calculated the attributable fraction of the temperature exposure in $2021$, in each of the $87$ Italian cities. The caterpillar plot in Figure \ref{fig:af_smooth_counterfactual} shows the estimated AF (red dots), together with $95\%$ credible intervals as well as the AF obtained from a counterfactual scenario where the deprivation index is equal to the median deprivation in all cities (blue triangles). It can be seen that for the top $9$ cities, which are the locations with the highest deprivation index, a significant difference between the true AF and the counterfactual AF is found.

\section{Conclusion}
While DLNMs are a popular technique to flexibly quantify the delayed effect of an exposure, the possibility of effect modification is often ignored. However, acknowledging it is crucial because it highlights meaningful differences between subgroups or contexts. In particular in a spatial context, area-level characteristics can modify the effect of the exposure on the response. In this paper, we extended the (spatial) framework introduced by \cite{Rutten2026} to allow for spatial effect modification. We allowed for a linear or a smooth interaction with a continuous covariate, offering great flexibility in capturing modifying effects. Although the method has been developed in a geographical context, the same methodology can easily be applied in a non-spatial context, studying for example effect modification across population subgroups such as different age groups.

The simulation study showed that the proposed extension performs well in terms of RMSE and coverage of the RRs. Moreover, the importance of penalization of the DLNM was illustrated, especially when allowing for a smooth interaction effect. In the presence of an effect modifier, the DLNM-LPS method ignoring the modifier consistently underperformed. Furthermore, our proposed method showed smaller RMSE than the meta-analysis method, especially for small areas. Accuracy of the model fitted in \texttt{bam} was similar to the LPS method but, by leveraging Bayesian P-splines, the LPS method showed superior computational speed. Moreover, the Bayesian framework offers the advantage of better quantification of uncertainty, leading to an increased coverage compared to \texttt{bam}. 

Hence, our proposed method offers several advantages over previously proposed approaches. In contrast to the meta-analysis method, our method enables information sharing between areas, leading to more robust estimates. Moreover, our method allows for penalization of the spline basis functions, which is not straightforward in commonly used Bayesian software like integrated nested Laplace approximation (INLA). Furthermore, the method performs consistently faster than the \texttt{bam} package and a Leroux structured random effect can easily be incorporated. 

There are some limitations to our approach though. First of all, similar to the empirical Bayesian method, the uncertainty of the hyperparameters is not taken into account. However, the DLNM-LPS method does not show undercoverage problems. Moreover, the method assumes one known effect modifier. Extending the model to allow for multiple effect modifiers is feasible in theory, but will become computationally highly demanding. Moreover, the assumption of the effect modifier being known, is not straightforward in practice. Extending the model to allow for randomly varying exposure-response relationship, is therefore a useful direction for future research.

In summary, we introduced a novel method to allow for spatial effect modification within the DLNM framework. By enabling information sharing, supporting penalization and incorporating spatial random effects while retaining computational efficiency, the approach addresses several key limitations of existing methods. These advantages make our proposed method well suited for small area data analyses where effect modification of complex exposure-lag-response relationships is of interest.

\section*{Supplementary materials}
\textbf{Online Appendix}: Contains additional tables and figures. (Supplementary.pdf)\\


\section*{Acknowledgements}
The computational resources and services used in this work were provided by the VSC (Flemish Supercomputer Center), funded by the Research Foundation - Flanders (FWO) and the Flemish Government - department EWI.

\section*{Funding}
T.N. gratefully acknowledges funding from Research Foundation - Flanders (grant no. G0A3M24N).
\section*{Competing Interest Statement}
The authors have declared no competing interest.

\newpage
\newpage
\bibliographystyle{apalike}
\bibliography{References}

\end{document}


\maketitle
\section{Matrix formulation model}
\subsection{Construction cross-basis}
The cross-basis function $ s(x_{t,j}, \ldots , x_{t-L,j} ; \boldsymbol{\theta_j})$ is constructed as:
\begin{align}
    s(x_{t,j}, \ldots , x_{t-L,j} ; \boldsymbol{\theta_j})&=\sum_{i=1}^{v_x} \sum_{k=1}^{v_l}\left(\sum_{l=0}^L\widetilde{b}_i(x_{t-l,j})\shortsmile{b}_k(l) \right)\theta_{j,ik},
\end{align}
Define $(L+1) \times v_x$ dimensional basis matrix $\widetilde{B}_{t,j}$ by applying the functions $\{\widetilde{b}_i\}_{i=1}^{v_x}$ to the vector $\boldsymbol{q_{t,j}} = (x_{t,j}, x_{t-1,j}, \ldots, x_{t-L,j})^\top$. Similarly, $(L+1) \times v_l$ dimensional matrix $\shortsmile{B}$ can be constructed by applying basis $\{\shortsmile{b}_k\}_{k=1}^{v_l}$ to $\boldsymbol{l} = (0,1,\ldots,L)^\top$.  Thus, the cross-basis function $s(x_{t,j}, \ldots , x_{t-L,j} ; \boldsymbol{\theta_j})$, can be written as $s(x_{t,j}, \ldots , x_{t-L,j} ; \boldsymbol{\theta_j}) = \left( \mathbf{1}^{\top}_{L + 1} \, A_{t,j} \right) \boldsymbol{\theta_j} = \boldsymbol{w_{t,j}}^\top\boldsymbol{\theta_j}$, with $\mathbf{1}$ a vector of ones and $\boldsymbol{w_{t,j}}$ derived from the matrix $A_{t,j} = \left( \widetilde{B}_{t,j} \otimes 1_{v_l}^\top \right) \odot \left( 1_{v_x}^\top \otimes \shortsmile{B} \right)$, where $\otimes$ and $\odot$ represent the Kronecker and Hadamard products, respectively. The full $TJ \times(v_x\times v_l)$ dimensional matrix $W$, called the cross-basis matrix, can then be constructed by defining $\boldsymbol{w_{t,j}}$ the $tj$-th row of $W$.

Regardless of the basis specification $\{\tilde{c}_r\}_{r=1}^{v_z}$, the function $s(x_{t,j}, \ldots , x_{t-L,j} ; \boldsymbol{\theta}(f(z_j)))$ can be written in vector notation as:
\begin{align*}
      s(x_{t,j}, \ldots , x_{t-L,j} ; \boldsymbol{\theta}(f(z_j))) = \boldsymbol{w_{t,j}}^\top \boldsymbol{\theta}^{(1)}+\boldsymbol{v_{t,j}}^\top \boldsymbol{\theta}^{(2)},
\end{align*}
with $\boldsymbol{w_{t,j}}$ and $\boldsymbol{v_{t,j}} = (\boldsymbol{w_{t,j}}^\top\tilde{c}_1(z_{j}), \ldots, \boldsymbol{w_{t,j}}^\top\tilde{c}_{v_z}(z_{j}))^\top$ the $tj$-th row from $TJ \times (v_xv_l)$ dimensional matrix $W$ and $TJ \times (v_xv_lv_z)$ dimensional matrix $V$ respectively.

\subsection{Construction design matrix DLNM-LPS with effect modification}
In the presence of an effect modifier, the main effect of $z$ should be included in the model. For a categorical effect modifier, the $TJ \times (F-1)$ dimensional design matrix $Z$ can be easily constructed by letting the $tj$-th row equal the dummy-variable representation of $z_j$. In case of a continuous effect modifier, the covariate can be included either linearly or through a smooth function $g$. Hence, for a linear effect, the design matrix $Z$ has dimensions $TJ \times 1$ where $z_{t,j} = z_j$. For a smooth main effect, a $v_{z_2}$ dimensional B-spline basis $\{\tilde{d}_q\}_{q=1}^{v_{z_2}}$ needs to be chosen, resulting in a $TJ \times v_{z_2}$ dimensional design matrix $Z$ with $\boldsymbol{z}_{t,j} = (\tilde{d}_1(z_j), \ldots,\tilde{d}_{v_{z_2}}(z_j))^\top$. Denote the associated coefficients by $\boldsymbol{\gamma}$. Let covariate vector $\boldsymbol{a_{t,j}} = (1, a_{t,j,1},\ldots a_{t,j,H})^\top$ be the $tj$-th row of $TJ \times (H+1)$ dimensional matrix $A$ and denote $\boldsymbol{\beta} = (\beta_0, \beta_1, \ldots \beta_H)^\top$. Finally, let $\boldsymbol{u} = (u_1, \ldots u_J)^\top$ be a vector of random effects. If we denote $\boldsymbol{y} = \{y_{t,j}: t=1,\dots,T; j=1,\dots,J\}$ and $\mathbb{E}(\boldsymbol{y}) = \boldsymbol{\mu}$, the model can be written in matrix notation as:
\[\log(\boldsymbol{\mu}) = A\boldsymbol{\beta} +Z\boldsymbol{\gamma}+ W\boldsymbol{\theta}^{(1)}+ V\boldsymbol{\theta}^{(2)} + M\boldsymbol{u} = H\boldsymbol{\xi},\]
where $H=[A:Z:W:V:M]$ with $M$ the design matrix of the random vector $\boldsymbol{u}$ and $\boldsymbol{\xi}=(\boldsymbol{\beta}^{\top},\boldsymbol{\gamma}^{\top},\boldsymbol{\theta}^{(1)\top},\boldsymbol{\theta}^{(2)\top}, \boldsymbol{u}^{\top})^{\top}$.

\section{Simulation set-up}
Consider the smooth function: 
\begin{align*}
    f_j\cdot w_j(x_{t-l,j},l,z_j) = 0.1 \times \sum_{p=1}^{5}\delta_{pj}(z_j)(x_{t-l,j}-5)^{p-1}\exp(-l/d_j(z_j))
\end{align*}
The definition of $\delta_{pj}(z_j)$ and $d_j(z_j)$ depends on the complexity of the effect modification (i.e. linear or complex). Similar to the simulation study of \cite{Gasparrini2017}, we denote by vector $\delta = (0.211881,0.1406585,-0.0982663,0.0153671,-0.0006265)^\top$ the baseline coefficients. Define in addition the vector $\Delta = (0.2,0.1,0.5,0.3,0.15)^\top$. In the first setting, a simple case of effect modification was constructed with effect modifying covariate $z_j \sim \mathcal{N}(0,0.4^2)$. Then, $\delta_{pj}(z_j) = \delta_p (1+\Delta_pz_j)$ for $p = 1,\dots,5$ and $j = 1,\dots,73$. Furthermore, we defined $d_j(z_j) = 2$. In the second setting, we considered a more complicated case of effect modification. The covariate $z_j$ was again simulated from $\mathcal{N}(0,0.4^2)$ but afterwards, transformed to the range $0$ to $2\pi$ such that we could define $\delta_{pj}(z_j) = \delta_p(1+\Delta_p\cos(z_j))$ and $d_j(z_j) = 2(1+0.1 \sin(z_j))$.

\section{Simulation metrics}
In each of the $250$ simulations and for each method, we estimated the RMSE of the lag-specific and overall cumulative RR. In every simulation, the lag-specific RR was calculated for every $x = 0,0.25,\ldots,9.75,10$, each lag $l=0,1,\ldots 8$ and each area $j = 1,\ldots,73$ with effect modifier $z_j$. Afterwards, the across-the-surface statistics were summarized as:
\begin{align*}
    \text{RMSE} = \sqrt{\frac{1}{ 73 \times 41 \times 9}\sum_{j=1}^{73}\sum_{x=0}^{10}\sum_{l=0}^{8}{\biggl(\frac{1}{250}\sum_{s = 1}^{250}\bigl(\log\bigl(\widehat{RR}^{s}_{x,x_0,z_j}(l)\bigl) - \log\bigl(RR^{\text{true}}_{x,x_0,z_j}(l)\bigl)\bigl)^2\biggl) }},
\end{align*}
for the lag-specific RR and 
\begin{align*}
    \text{RMSE} = \sqrt{\frac{1}{ 73 \times 41 \times 9}\sum_{j=1}^{73}\sum_{x=0}^{10}{\biggl(\frac{1}{250}\sum_{s = 1}^{250}\bigl(\log\bigl(\widehat{RR}^{\text{overall},s}_{x,x_0,z_j}\bigl) - \log\bigl(RR^{\text{overall,true}}_{x,x_0,z_j}\bigl)\bigl)^2\biggl) }},
\end{align*}
for the overall cumulative RR. Furthermore, we calculated the coverage of the $95\%$ credible intervals using the posterior distribution of each evaluated feature, resulting in:
\begin{align*}
    \text{coverage} = \frac{1}{ 73 \times 41 \times 9}\sum_{j=1}^{73}\sum_{x=0}^{10}\sum_{l=0}^{8}{\biggl(\frac{1}{250}\sum_{s = 1}^{250}\bigl( \log\bigl(RR^{\text{true}}_{x,x_0,z_j}(l)\bigl) \in \text{CI}_s\bigl)\biggl) },
\end{align*}
and
\begin{align*}
    \text{coverage} = \frac{1}{ 73 \times 41 \times 9}\sum_{j=1}^{73}\sum_{x=0}^{10}\sum_{l=0}^{8}{\biggl(\frac{1}{250}\sum_{s = 1}^{250}I\bigl(\log\bigl(RR^{\text{true}}_{x,x_0,z_j}(l)\bigl ) \in \text{CI}_s\bigl)\biggl) },
\end{align*}
where $\text{CI}_s$ is the relevant credible interval, calculated from simulation $s$.
\section{Additional Tables}

\begin{table}[h!]
\caption{Simulation results (negative binomial model) for linear effect modification: RMSE of lag-specific RR (RMSE RR), RMSE of cumulative RR, over lag $0-8$, (RMSE RR overall), coverage of lag-specific RR (cov RR) and coverage of cumulative RR (cov RR overall). The computation time is reported in seconds (averaged over $10$ runs), as well as the percentage of failed (or timed-out) simulations.}
\label{tab:simulation_study_linear_NB}
\centering
\begin{adjustbox}{width=\linewidth} 
\begin{tabular}{lllllcccccc}
\hline
 \textbf{Area} &  \textbf{Method} & \textbf{Modification} &\textbf{Penalty} & \textbf{RE}& \textbf{failed} & \textbf{time} & \textbf{cov RR} &\textbf{cov RR} & \textbf{RMSE RR} &\textbf{RMSE RR}  \\ 
  &   &  &  & &  &  &  &\textbf{overall} &  &\textbf{overall}  \\ \hline
 \multirow{11}{*}{Small} &  Laplace (LPS)& smooth & yes & Leroux  & $0$ & $107.31$ & $0.9768$ & $0.9666$ & $0.0174$ & $0.0753$ \\
                          &  Laplace (LPS)& linear & yes & Leroux & $0$ & $39.30$ & $0.9754$ & $0.9617$ & $0.0161$ & $0.0685$  \\
                          &  Laplace (LPS) & binary & yes & Leroux & $0$ & $32.88$ & $0.9715$ & $0.9295$ & $0.0173$ & $0.0758$ \\
                          &   Laplace (LPS) & none & yes & Leroux & $0$ & $23.69$ & $0.9529$ & $0.8251$ & $0.0200$ & $0.0875$  \\
                          &  Laplace & smooth & no & Leroux & $0$ & $29.04$ & $0.9401$ & $0.9452$ & $0.0681$ & $0.1816$ \\
                          &  Laplace & linear & no & Leroux & $0$ & $11.21$ & $0.9400$ & $0.9340$ & $0.0280$ & $0.0871$ \\                          
                          &  BAM & smooth & yes & ind. & $0.22$ & $387.86$ & $0.8729$ & $0.8405$ & $0.0111$ & $0.0558$ \\
                          &  BAM & linear & yes & ind. & $0$ & $39.50$ & $0.8242$ & $0.7405$ & $0.0115$ & $0.0597$ \\
                          &  CTS & linear & no & $\cdot$ & $0$ & $0.37$ & $0.9669$ & $0.9610$ & $0.0304$ &  $0.0951$ \\
                         &  Meta &  smooth & no & $\cdot$ & $0$ & $39.13$ & $0.9722$ & $0.9701$ & $0.0641$ & $0.2036$ \\
                          &  Meta & linear & no & $\cdot$ &$0$ & $31.99$ & $0.9737$ & $0.9707$ & $0.0412$ & $0.1332$ \\

\hline
  \multirow{11}{*}{Large} &  Laplace (LPS) & smooth & yes & Leroux & $0$ & $95.21$ & $0.9700$ & $0.9615$ & $0.0104$ & $0.0442$ \\
                          &  Laplace (LPS) & linear & yes & Leroux & $0$ & $34.54$ & $0.9671$ & $0.9576$ & $0.0100$ &  $0.0411$  \\
                          &  Laplace (LPS) & binary & yes & Leroux & $0$ & $27.23$ & $0.9480$ & $0.8182$ & $0.0109$ & $0.0480$ \\
                          & Laplace (LPS)& none & yes & Leroux & $0$ & $20.28$ & $0.9000$ & $0.5920$ & $0.0113$ & $0.0531$ \\
                          &  Laplace & smooth & no & Leroux & $0$ & $17.07$ & $0.9387$ & $0.9291$ & $0.0180$ & $0.0571$ \\
                          &  Laplace & linear & no & Leroux & $0$ & $8.72$ & $0.9394$ & $0.9228$ & $0.0104$ & $0.0355$ \\
                          &  BAM & smooth & yes & ind. & $0.23$ & $414.79$ & $0.9249$ & $0.8897$ & $0.0076$ & $0.0362$ \\
                          &  BAM & linear & yes & ind. & $0$ & $46.08$ & $0.8940$ & $0.8432$ & $0.0076$  & $0.0380$ \\
                          &  CTS & linear & no & $\cdot$ & $0$ & $0.39$ & $0.9461$ & $0.9306$ & $0.0104$ & $0.0356$ \\
                          &  Meta &  smooth & no & $\cdot$ & $0$ & $40.42$ & $0.9714$ & $0.9700$& $0.0226$  & $0.0707$ \\
                          &  Meta & linear & no & $\cdot$ & $0$ & $31.87$ & $0.9745$ & $0.9725$ & $0.0187$  & $0.0589$ \\

\hline

\end{tabular}
\end{adjustbox}
\end{table}

\begin{table}[h!]
\caption{Simulation results (negative binomial model) for smooth effect modification: RMSE of lag-specific RR (RMSE RR), RMSE of cumulative RR, over lag $0-8$, (RMSE RR overall), coverage of lag-specific RR (cov RR) and coverage of cumulative RR (cov RR overall). The computation time is reported in seconds (averaged over $10$ runs), as well as the percentage of failed (or timed-out) simulations.}
\label{tab:simulation_study_smooth_NB}
\centering
\begin{adjustbox}{width=\linewidth} 
\begin{tabular}{lllllcccccc}
\hline
 \textbf{Area} &  \textbf{Method} & \textbf{Modification} &\textbf{Penalty} & \textbf{RE}& \textbf{failed} & \textbf{time} & \textbf{cov RR} &\textbf{cov RR} & \textbf{RMSE RR} &\textbf{RMSE RR}  \\ 
  &   &  &  & &  &  &  &\textbf{overall} &  &\textbf{overall}  \\ \hline
 \multirow{11}{*}{Small} &  Laplace (LPS) & smooth & yes & Leroux & $0$ & $110.75$ & $0.9828$ & $0.9650$ & $0.0193$ & $0.0906$ \\
                          &  Laplace (LPS) & linear & yes & Leroux & $0$ & $40.53$ & $0.9617$ & $0.9582$ & $0.0380$ & $0.1738$  \\
                         &   Laplace (LPS)& binary & yes & Leroux & $0$ & $36.18$ & $0.9511$ & $0.7816$ & $0.0212$ & $0.1073$ \\
                          &  Laplace (LPS) & none & yes & Leroux & $0$ & $24.62$ & $0.9339$ & $0.6837$ & $0.0226$ & $0.1066$ \\
                          &  Laplace & smooth & no & Leroux & $0$ & $38.87$ & $0.9455$ & $0.9448$ & $0.2020$ & $0.8508$ \\
                          &  Laplace & linear & no & Leroux & $0$ & $12.07$ & $0.9218$ & $0.7797$ & $0.0388$ & $0.1426$ \\
                          & BAM & smooth & yes & ind. & $0.16$ & $380.68$ & $0.9258$ & $0.8830$ & $0.0131$ & $0.0664$ \\
                          &  BAM & linear & yes & ind. & $0$ & $297.02$ & $0.8123$ & $0.6263$ & $0.0195$ & $0.1049$ \\
                          &  CTS & linear & no & $\cdot$ & $0$ & $0.52$ & $0.9633$ & $0.8753$ & $0.0403$ & $0.1448$ \\
                          &  Meta &  smooth & no & $\cdot$ & $0$ & $44.06$ & $0.9744$ & $0.9671$ & $0.1327$ & $0.4427$ \\
                          &  Meta & linear & no & $\cdot$ & $0$ & $37.63$ & $0.9714$ & $0.9287$ & $0.0510$ & $0.1958$ \\

\hline
 \multirow{11}{*}{Large} &  Laplace (LPS) & smooth & yes & Leroux  & $0$ & $90.47$ & $0.9739$ & $0.9597$ & $0.0113$ & $0.0484$ \\
                          &  Laplace (LPS) & linear & yes & Leroux  & $0$ & $33.95$ & $0.9037$ & $0.5498$ & $0.0180$ & $0.0825$  \\
                          &  Laplace (LPS) & binary & yes & Leroux  & $0$ & $29.03$ & $0.8867$ & $0.5619$ & $0.0137$ & $0.0723$ \\
                          &  Laplace (LPS) & none & yes & Leroux &  $0$ & $19.45$ & $0.8550$ & $0.4337$ & $0.0142$ & $0.0738$ \\
                          &  Laplace & smooth & no & Leroux  & $0$ & $19.18$ & $0.9465$ & $0.9430$ & $0.0201$ & $0.0611$\\
                          &  Laplace & linear & no & Leroux  & $0$ & $7.45$ & $0.8029$ & $0.4021$ & $0.0150$ & $0.0716$  \\
                          & BAM & smooth & yes & ind. & $0.14$ & $348.37$ & $0.9373$ & $0.9169$ & $0.0080$ & $0.0387$ \\
                          &  BAM & linear & yes & ind.  & $0$ & $53.52$ & $0.7048$ & $0.3017$ & $0.0121$ & $0.0687$ \\
                          &  CTS & linear & no & $\cdot$ & $0$ & $0.37$ & $0.8156$ & $0.4465$ & $0.0150$ & $0.0715$ \\
                          &  Meta &  smooth & no & $\cdot$ & $0$ & $44.05$ & $0.9776$ & $0.9760$ & $0.0239$ & $0.0757$ \\
                          &  Meta & linear & no & $\cdot$ & $0$ & $34.01$ & $0.9734$ & $0.9626$ & $0.0203$ & $0.0690$ \\

\hline

\end{tabular}
\end{adjustbox}
\end{table}

\section{Additional Figures}
\begin{figure}[H]
    \centering
        \includegraphics[width=\linewidth]{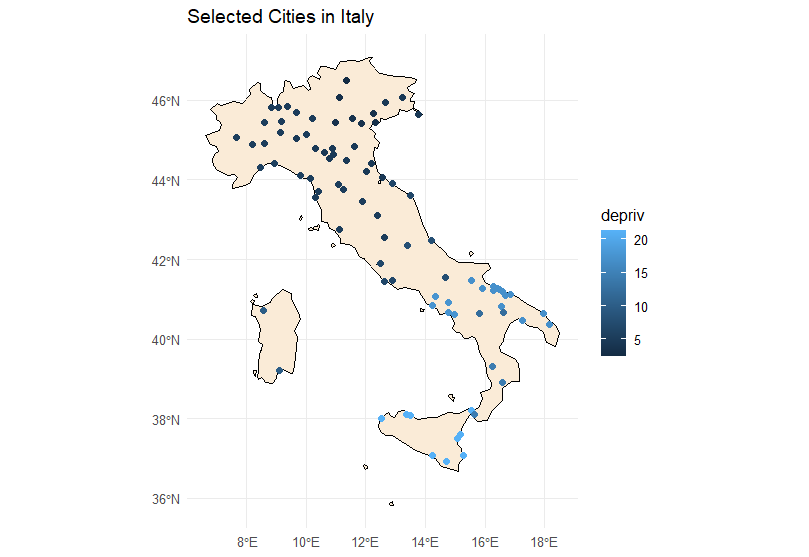}
        \caption{Map of selected cities in Italy, coloured by the deprivation index.}
        \label{fig:map_italy}
\end{figure}

\begin{figure}[H]
    \centering
    \includegraphics[width=0.7\linewidth]{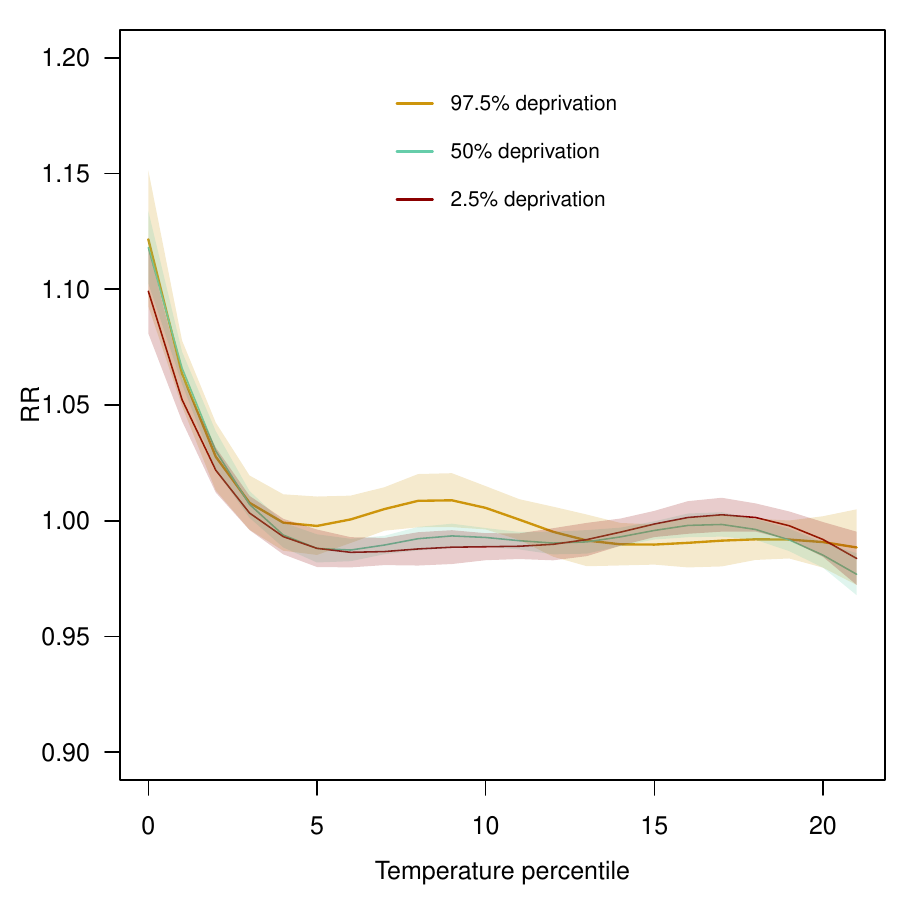}
    \caption{Lag-response relationship for a $0.95\%$ temperature quantile, compared to a reference temperature quantile of $0.60\%$, for $3$ deprivation levels.}
    \label{fig:RR_lag_specific}
\end{figure}

\begin{figure}[H]
    \centering
    \includegraphics[width=0.7\linewidth]{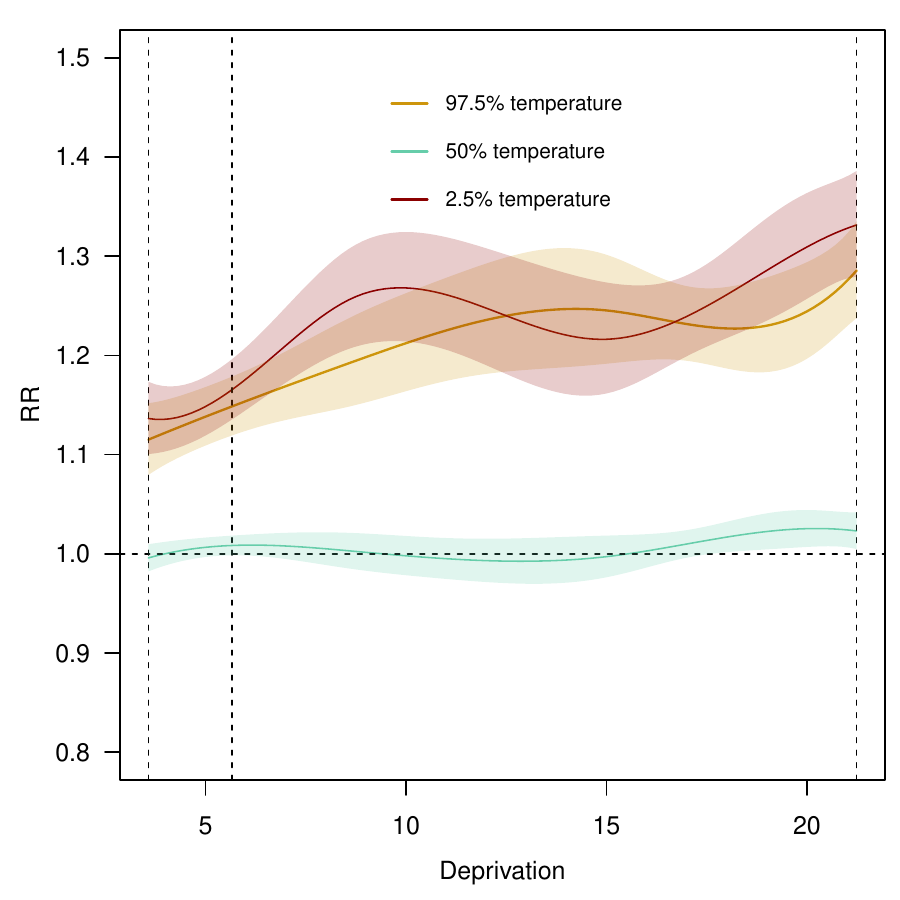}
    \caption{Overall RR, compared to a temperature percentile of $0.6$ and cumulated over 21 days, versus the deprivation index for three different temperature percentiles. The vertical lines represent the $2.5\%, 50\%$ and $97.5\%$ quantiles of the deprivation index. The shaded bands indicate the $95\%$ credible intervals.}
    \label{fig:RR_by_cov_smooth}
\end{figure}

\begin{figure}[H]
    \centering
        \includegraphics[width=0.7\linewidth]{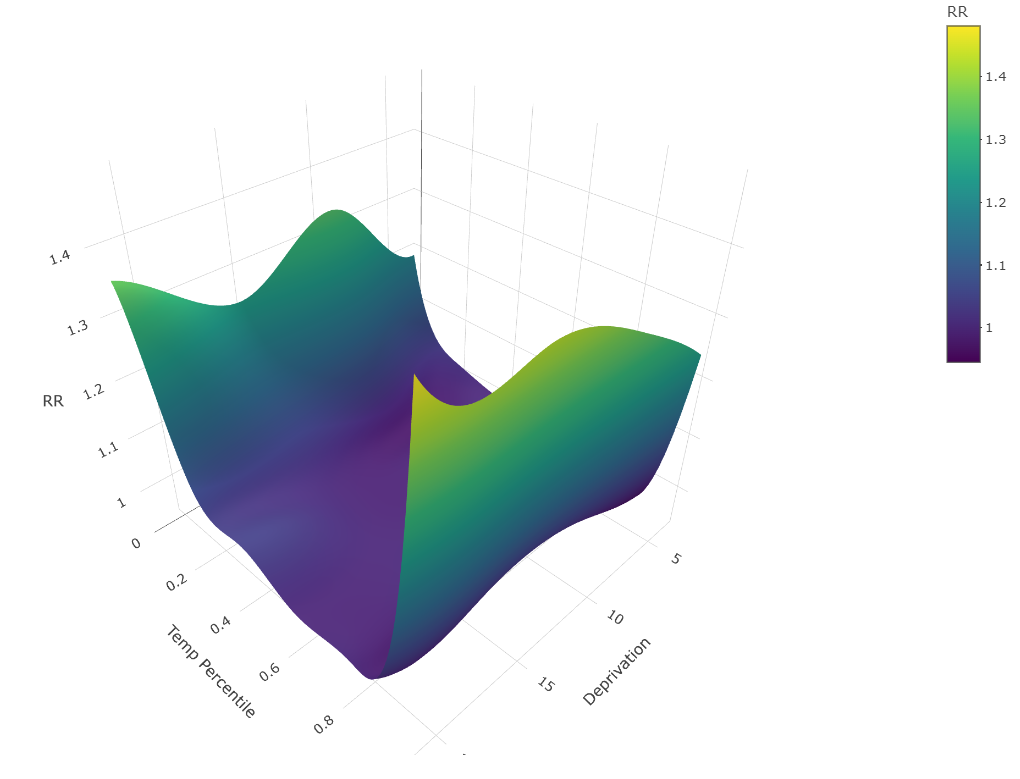}
        \caption{3D plot showing the overall estimated RR, cumulated over $21$ days and compared to a reference temperature percentile of $0.6$, for different temperature percentiles and deprivation indices.}
        \label{fig:3d_italy}
\end{figure}

\begin{figure}[H]
    \centering
    \includegraphics[width=1\linewidth]{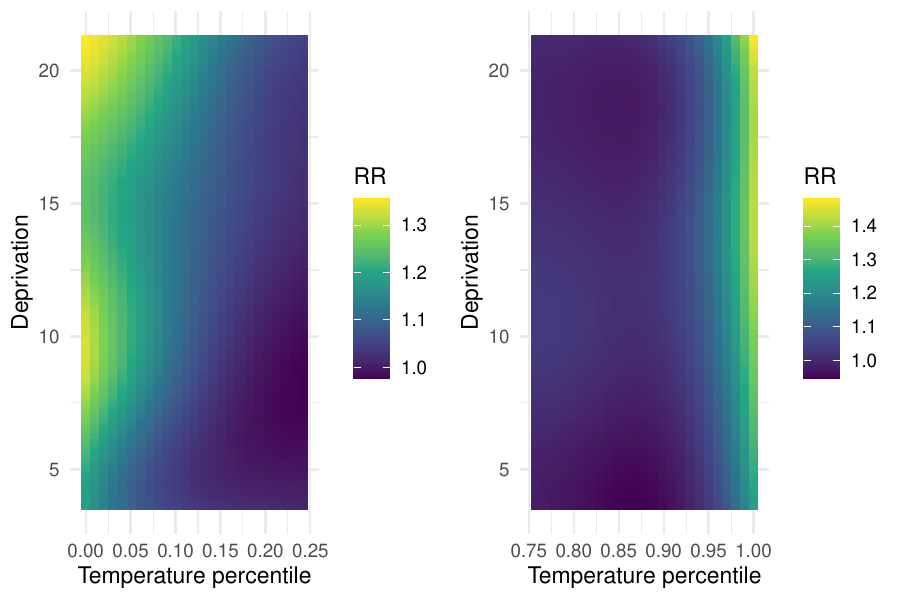}
    \caption{Contour plot showing the overall estimated RR, cumulated over $21$ days and compared to a reference temperature percentile of $0.6$, for different temperature percentiles and deprivation indices. The left figure shows the relationship for low temperature percentiles while the right figure shows the relationship for high temperature percentiles.}
    \label{fig:placeholder}
\end{figure}

\begin{figure}[H]
    \centering
        \includegraphics[width=0.7\linewidth]{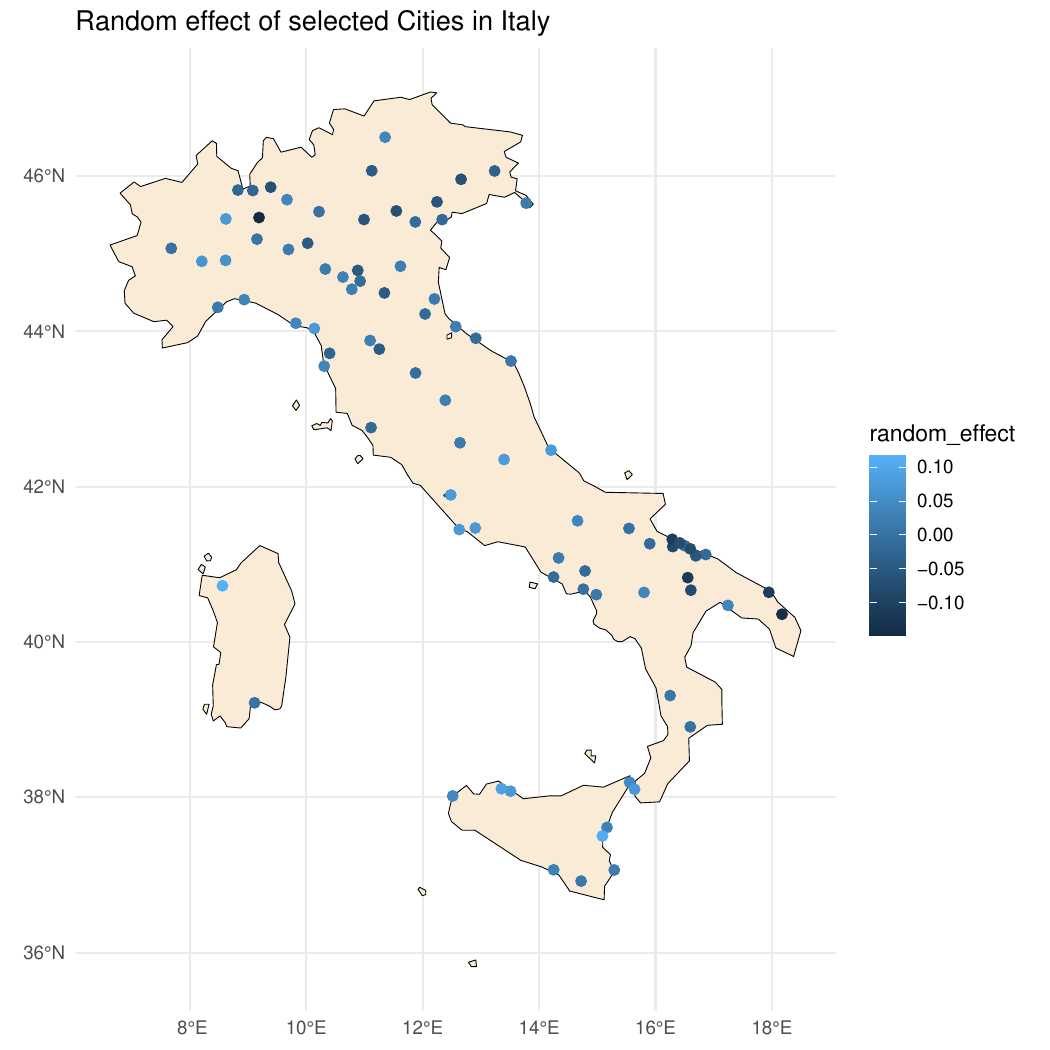}
        \caption{Estimated random effects for the $83$ selected Italian cities.}
        \label{fig:map_italy_re}
\end{figure}

\begin{figure}[H]
    \centering
        \includegraphics[width=0.7\linewidth]{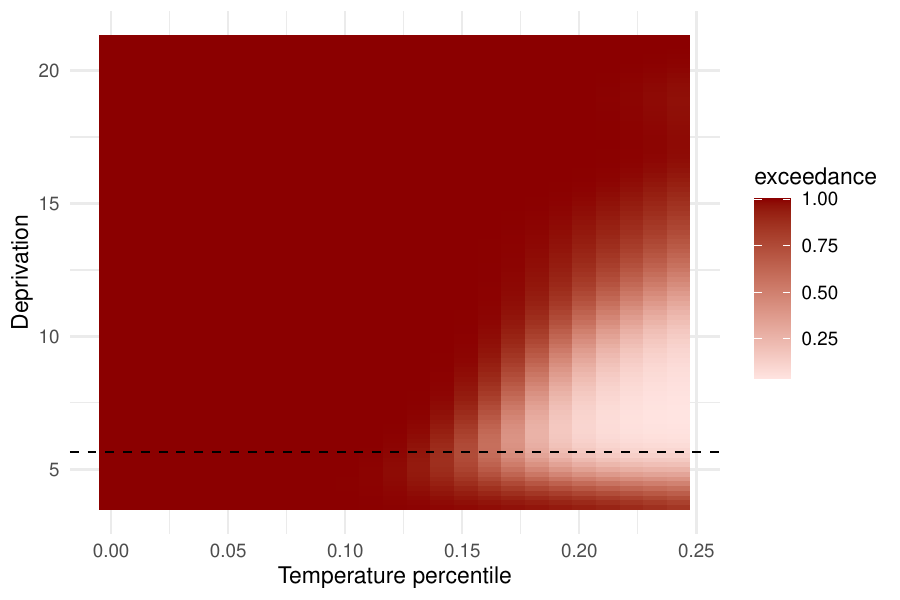}
        \caption{Posterior probability of a RR exceeding $1$, for different (low) temperature percentiles and deprivation indices. The dashed horizontal line represents the median deprivation index.}
        \label{fig:exceedance_prob_italy_low}
\end{figure}

\begin{figure}[H]
    \centering
        \includegraphics[width=0.7\linewidth]{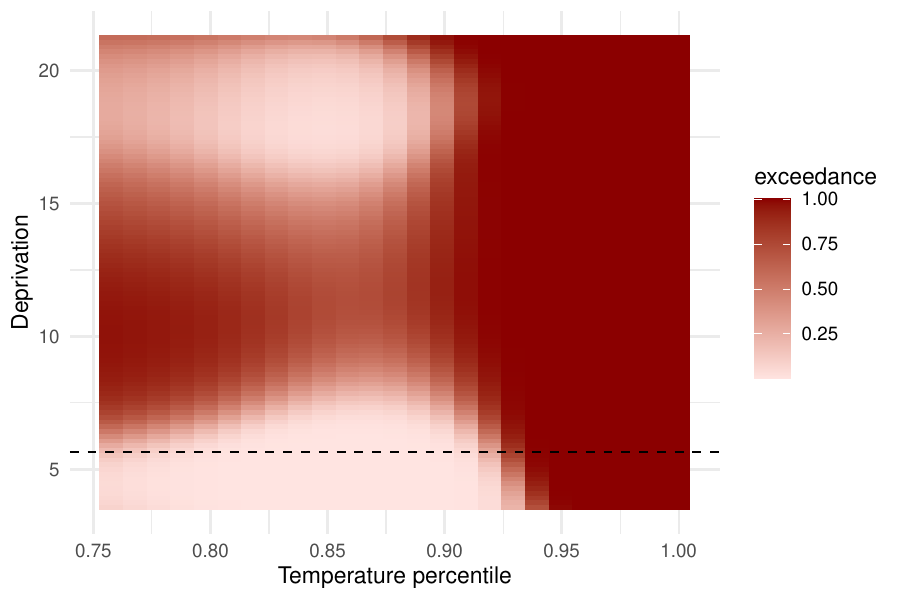}
        \caption{Posterior probability of a RR exceeding $1$, for different (high) temperature percentiles and deprivation indices. The dashed horizontal line represents the median deprivation index.}
        \label{fig:exceedance_prob_italy_high}
\end{figure}

\newpage
\bibliographystyle{apalike}
\bibliography{References}